\documentclass[preprint,a4paper,notoc]{JHEP3}
\usepackage{amsmath}
\usepackage{epsfig}
\usepackage{multicol}
\usepackage{psfrag}
\usepackage{amssymb}
\usepackage{bbm}
\usepackage{cite}
\usepackage{subfigure}

\newcommand{\parbreak}{\vskip 0.12cm}
\newcommand{\Eq}[1]{{Eq.~(\ref{#1})}}

\newcommand{\ZA}{{Z_{\rm\scriptscriptstyle A}}}
\newcommand{\cA}{{c_{\rm\scriptscriptstyle A}}}
\newcommand{\bA}{{b_{\rm\scriptscriptstyle A}}}
\newcommand{\ZVstat}{{Z_{\rm\scriptscriptstyle V}^{{\rm stat}}}}
\newcommand{\ZAstat}{{Z_{\rm\scriptscriptstyle A}^{{\rm stat}}}}
\newcommand{\cVstat}{{c_{\rm\scriptscriptstyle V}^{{\rm stat}}}}
\newcommand{\cAstat}{{c_{\rm\scriptscriptstyle A}^{{\rm stat}}}}
\newcommand{\cVstatone}{{c_{\rm\scriptscriptstyle V}^{{\rm stat}(1)}}}

\newcommand{\cVstatEH}{{c_{\rm\scriptscriptstyle V,EH}^{{\rm stat}}}}
\newcommand{\cVstatAPE}{{c_{\rm\scriptscriptstyle V,APE}^{{\rm stat}}}}
\newcommand{\cVstatHYPone}{{c_{\rm\scriptscriptstyle V,HYP1}^{{\rm stat}}}}
\newcommand{\cVstatHYPtwo}{{c_{\rm\scriptscriptstyle V,HYP2}^{{\rm stat}}}}
\newcommand{\cVstatoneEH}{{c_{\rm\scriptscriptstyle V,EH}^{{\rm stat}(1)}}}
\newcommand{\cVstatoneAPE}{{c_{\rm\scriptscriptstyle V,APE}^{{\rm stat}(1)}}}

\newcommand{\bVstatoneEH}{{b_{\rm\scriptscriptstyle V,EH}^{{\rm stat}(1)}}}
\newcommand{\bVstatoneAPE}{{b_{\rm\scriptscriptstyle V,APE}^{{\rm stat}(1)}}}

\newcommand{\bVstat}{{b_{\rm\scriptscriptstyle V}^{{\rm stat}}}}
\newcommand{\bAstat}{{b_{\rm\scriptscriptstyle A}^{{\rm stat}}}}

\newcommand{\bVstatone}{{b_{\rm\scriptscriptstyle V}^{{\rm stat}(1)}}}
\newcommand{\bVstatEH}{{b_{\rm\scriptscriptstyle V,EH}^{{\rm stat}}}}
\newcommand{\bVstatAPE}{{b_{\rm\scriptscriptstyle V,APE}^{{\rm stat}}}}

\newcommand{\ZAZVEH}{\left[{Z_{\rm\scriptscriptstyle A}^{{\rm stat}}}/{Z_{\rm\scriptscriptstyle V}^{{\rm stat}}}\right]_{\rm\scriptscriptstyle EH}}
\newcommand{\ZAZVAPE}{\left[{Z_{\rm\scriptscriptstyle A}^{{\rm stat}}}/{Z_{\rm\scriptscriptstyle V}^{{\rm stat}}}\right]_{\rm\scriptscriptstyle APE}}
\newcommand{\ZAZVHYPone}{\left[{Z_{\rm\scriptscriptstyle A}^{{\rm stat}}}/{Z_{\rm\scriptscriptstyle V}^{{\rm stat}}}\right]_{\rm\scriptscriptstyle HYP1}}
\newcommand{\ZAZVHYPtwo}{\left[{Z_{\rm\scriptscriptstyle A}^{{\rm stat}}}/{Z_{\rm\scriptscriptstyle V}^{{\rm stat}}}\right]_{\rm\scriptscriptstyle HYP2}}
\newcommand{\Astat}{{A_{0}^{\rm stat}}}

\newcommand{\Vstat}{{V_{0}^{\rm stat}}}
\newcommand{\Vkstat}{{V_{k}^{\rm stat}}}
\newcommand{\rA}{{\rm\scriptscriptstyle A}}
\newcommand{\rV}{{\rm\scriptscriptstyle V}}
\newcommand{\rVA}{{\rm\scriptscriptstyle VA}}
\newcommand{\rVP}{{\rm\scriptscriptstyle VP}}

\newcommand{\ct}{{c_{\rm t}}}
\newcommand{\cttilde}{{\tilde c_{\rm t}}}
\newcommand{\mq}{{m_{\rm q}}}
\newcommand{\mcr}{{m_{\rm cr}}}

\title{Non-perturbative renormalization of the static vector current and its ${\rm O}(a)$-improvement in quenched QCD}

\author{\epsfig{figure=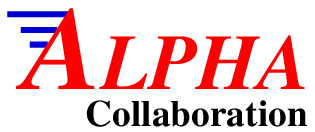,width=2.5cm}}

\author{
  Filippo Palombi\\
  DESY, Theory Group, Platanenallee 6, D-15738 Zeuthen, Germany\\
  E-mail: \email{filippo.palombi@desy.de}}

\preprint{SFB/CPP-07-24 \\ DESY 07-079 \\ June 2007}

\abstract{We carry out the renormalization and the Symanzik 
${\rm O}(a)$-improvement programme for the static vector current 
in quenched lattice QCD. The scale independent ratio of the 
renormalization constants of the static vector and axial currents 
is obtained non-perturbatively from an axial Ward identity with 
Wilson-type light quarks and various lattice discretizations of 
the static action. The improvement coefficients $\cVstat$ and 
$\bVstat$ are obtained up to ${\rm O}(g_0^4)$-terms by enforcing 
improvement conditions respectively on the axial Ward identity 
and a three-point correlator of the static vector current. A 
comparison between the non-perturbative estimates and the 
corresponding one-loop results shows a non-negligible effect of 
the ${\rm O}(g_0^4)$-terms on the improvement coefficients but a 
good accuracy of the perturbative description of the ratio of 
the renormalization constants.}

\keywords{HQET, non-perturbative renormalization, lattice QCD}

\newcommand{\bx}{{\mathbf{x}}}
\newcommand{\by}{{\mathbf{y}}}
\newcommand{\bz}{{\mathbf{z}}}

\newcommand{\cO}{{\cal O}}

\newcommand{\cR}{{\cal R}}

\newcommand{\cT}{{\cal T}}

\newcommand\csw{c_{\rm sw}}

\newcommand{\Dslash}{\relax{\kern+.25em / \kern-.70em D}}

\newcommand{\rh}{{\rm h}}

\newcommand{\rqq}{{\rm q}}
\newcommand{\rI}{{\rm I}}

\begin{document}


\section{Introduction}

Semileptonic decays of $B$-mesons constitute a very important source
of experimental information in $B$-physics. They  have been and are
currently being investigated as a part of the research programmes of
BaBar \cite{Aubert:2006ry} and CLEO \cite{Adam:2007pv}. The
prototype for such decays is $B^0\to \pi^-\ell^+\nu$. Once the
amplitude of this process is known, an experimental measurement of
its branching ratio allows in principle to extract the CKM matrix
element $|V_{ub}|$. From a theoretical point of view, the transition
is mediated by the heavy-light vector current, and the problem of
knowing the decay amplitude amounts to calculating the QCD matrix
element
\begin{equation}
\label{matrixelement} \langle \pi(p)|V_{\mu}|B(k)\rangle = \left(k +
p - q\frac{m_B^2 - m_\pi^2}{q^2}\right)_\mu f_+(q^2) +
\frac{m_B^2-m_\pi^2}{q^2}q_\mu f_0(q^2)\ ,
\end{equation}
or equivalently the form factors $f_{+/0}(q^2)$, with $q=k-p$ the
4-momentum transferred from the $B$-meson to the pion.
\parbreak
Given the large mass of the $b$-quark, a direct lattice calculation
of \Eq{matrixelement} requires tiny lattice spacings ($a\ll 1/(5 {\rm
GeV}))$ and big volumes ($L > 1.5\ {\rm fm}$) in order to
correctly reproduce the quark dynamics without squeezing the
$B$-meson at reasonably small light-quark masses. Various solutions
have been proposed to overcome this difficulty: the reader is
referred to \cite{Okamoto:2005zg,Onogi:2006km} for recent reviews.
Among these we mention the Heavy Quark Effective Theory (HQET) and
the Step Scaling Method (SSM).
\parbreak
In HQET \Eq{matrixelement} is expanded in inverse powers of the
$b$-quark mass. The leading contribution, also known as the static
approximation, describes the heavy quark in terms of a
renormalizable effective field theory. Although lattice simulations
in the original formulation \cite{Eichten:1989zv} were hampered by
large statistical fluctuations due to self-energy effects of the
heavy propagator, thanks to the recent introduction of new lattice
regularizations \cite{DellaMorte:2005yc} it is now possible to
simulate static quarks with much improved numerical precision.
\parbreak
The SSM, a relativistic technique based on finite size scaling, has
been proposed some years ago by the {\it Tor Vergata group} in
relation to a study of the heavy-light decay
constants~\cite{deDivitiis:2003wy} and meson masses
\cite{deDivitiis:2003iy}. It has been recently shown in
\cite{Guazzini:2006bn} that combining the SSM with HQET enables a
strict control of the mass extrapolations and a consequent
reduction of the corresponding systematic uncertainties.
\parbreak
From this point of view it would be of considerable interest to extend 
the combined approach ``HQET + SSM'' to \Eq{matrixelement}, since a 
first attempt to apply the {\it Tor Vergata} method to the form 
factors has been recently presented in \cite{Tantalo:2007ai}. The 
goal is ambitious in that observables such as \Eq{matrixelement} are 
intrinsically more complex than a decay constant or a meson 
mass, owing to the appearance of an additional mass scale to be 
identified with~$q^2$.
\parbreak
In this paper we concentrate on HQET. In view of a non-perturbative
computation of \Eq{matrixelement}, the static vector current must 
first be non-perturbatively renormalized. This task has been partially
accomplished, since in the static approximation the spatial
components $\Vkstat$ are renormalized by the same renormalization
constant $\ZAstat$ as the temporal component of the static axial
current $\Astat$. The Renormalization Group (RG) running of the
latter has been computed both in the quenched approximation
\cite{Heitger:2003xg} and with two dynamical quarks
\cite{DellaMorte:2006sv}. In order to compute the renormalization
constant $\ZVstat$ of the temporal component of the static vector
current $\Vstat$, we derive an axial Ward identity (WI), much in the
spirit of \cite{Aoki:1999ij,Hashimoto:2002pe}, relating $\ZVstat$ to
$\ZAstat$. The scale independent ratio $\ZAstat/\ZVstat$ is then
computed through an explicit implementation of the WI in the
Schr\"odinger functional (SF) at the chiral point. On-shell ${\rm
O}(a)$-improvement at zero light-quark mass is obtained by adding a
single counter-term to the static vector current, proportional to
the improvement coefficient $\cVstat$, which is then tuned according
to the request that the axial WI be satisfied at finite lattice
spacing up to ${\rm O}(a^2)$-terms.
\parbreak
The improvement of the static vector current $\Vstat$ at non-zero
light-quark mass, realized in principle through the introduction of
a second improvement coefficient $\bVstat$, is not easily achievable
in terms of the WI, which takes its simplest form in the chiral
limit. For this reason, we adopt a different improvement condition,
i.e we obtain $\bVstat$ by imposing that the ratio of a three-point
SF correlator of the static vector current at zero and non-zero
light-quark mass be the same in two different static regularizations
up to ${\rm O}(a^2)$-terms, thus determining the difference
$\Delta\bVstat$ corresponding to the chosen actions. This procedure
repeats the one adopted in \cite{DellaMorte:2005yc} for the
determination of $\bAstat$. In order to isolate the value of
$\bVstat$ corresponding to all the static discretizations, the
knowledge of $\bVstat$ is required for at least one of them. This is
a  difficult problem, which we solve only approximately by computing
$\bVstat$ at one-loop order in perturbation theory for the static
actions with the simplest lattice Feynman rules, i.e. the
Eichten-Hill (EH) and the APE ones. This somewhat unsatisfactory
solution introduces ${\rm O}(g_0^4)$ systematic uncertainties, which
are discussed in detail.
\parbreak
Other appealing applications where the static vector current plays a
r\^ole can be found within the domain of twisted mass QCD
\cite{Frezzotti:2000nk,Frezzotti:2003ni}, where the static axial
current acquires a vector component after a twist rotation of the
light-quark fields. This is the case, for instance, with the
computation of $B_B^{\rm stat}$, for which the matrix elements of the
$\Delta B=2$ four-fermion operators have to be normalized by
appropriate bilinear correlators of the static axial
current\cite{Palombi:inprep}.
\parbreak
The paper is organized as follows. The axial WI is derived in
sect.~2, where the notation is also established. Its implementation
in the framework of the SF is discussed in sect.~3. Sect.~4 is
devoted to a one-loop perturbative analysis of the lattice artefacts
in various WI topologies. In sect.~5 we present our non-perturbative
results for the improvement coefficient $\cVstat$ and the ${\rm
O}(a)$-improved ratio $\ZAstat/\ZVstat$, and in sect.~6 we discuss
the improvement coefficient $\bVstat$. Conclusions are drawn in
sect.~7. Additional tables containing perturbative and
non-perturbative results have been collected in appendix A.


\section{Formal derivation of the axial WI}

As for the theoretical derivation of the axial WI, we follow the
approach of \cite{Luscher:1998pe,Sommer:2006sj}. For the moment no
attention is paid to the specific regularization of the theory. We
assume a fermion content with an isospin doublet of degenerate
light-quarks $\psi^T=(\psi_1,\psi_2)$ and a single heavy-quark,
described by a pair of static fields $(\psi_\rh,\psi_{\bar \rh})$.
In order to set up the notation, we introduce the light-quark
isovector axial and vector currents and the pseudoscalar density
\begin{align}
A_\mu^a(x) & = \bar\psi(x)\gamma_\mu\gamma_5\frac{1}{2}\tau^a\psi(x)\ , \ \\
V_\mu^a(x) & = \bar\psi(x)\gamma_\mu\frac{1}{2}\tau^a\psi(x)\ , \ \\
P^a(x) & = \bar\psi(x)\gamma_5\frac{1}{2}\tau^a\psi(x)\ ,
\end{align}
as well as their heavy-light companions (for which we explicitly
indicate light-quark flavour indices)
\begin{align}
\hskip 1.2cm A_\mu^{k\rh}(x) & = \bar\psi_k(x)\gamma_\mu\gamma_5\psi_\rh(x), \label{hlaxial}  \\[1.7ex]
V_\mu^{k\rh}(x) & = \bar\psi_k(x)\gamma_\mu\psi_\rh(x), \label{hlvector} \\[1.7ex]
P^{\rh k}(x) & = \bar\psi_\rh(x)\gamma_5\psi_k(x)\ , \qquad k=1,2 \
. \label{hlpseudo}
\end{align}

The general WI follows from the invariance of the path integral
representation of correlation functions under chiral rotations of
the light-quark fields. In particular, we consider an axial
variation
\begin{equation}
\label{variation} \delta^a_{\rm\scriptscriptstyle A}\psi(x) =
\omega^a(x)\frac{1}{2}\tau^a\gamma_5\psi(x)\ , \qquad
\delta^a_{\rm\scriptscriptstyle A}\bar \psi(x) =
\omega^a(x)\bar\psi(x)\gamma_5\frac{1}{2}\tau^a\ ,
\end{equation}
where $\tau^a$ denotes a Pauli matrix acting on the isospin space
and $\omega^a(x)$ is a smooth function which vanishes outside some
bounded region $R$. Since the Pauli matrices are traceless, the
functional integration measure is invariant under \Eq{variation} and
we conclude that the correlation function of a given operator $\cO$
satisfies the equation
\begin{equation}
\label{genWI} \langle\cO\delta^a_{\rm\scriptscriptstyle A}S\rangle =
\langle \delta^a_{\rm\scriptscriptstyle A}\cO\rangle\ ,
\end{equation}
where
\begin{equation}
\delta^a_{\rm\scriptscriptstyle A}S = \int_R{\rm d}^4x\
\omega^a(x)\left\{-\partial_\mu A_\mu^a(x) + 2mP^a(x)\right\}
\end{equation}
represents the axial variation of the light-quark action and $m$
denotes the PCAC quark mass. We assume in what follows that $\cO$
factorizes into the product of two operators $\cO_{\rm int}$ and
$\cO_{\rm ext}$, polynomials in the basic fields and localized in
the interior and exterior of $R$ respectively. Accordingly,
\Eq{genWI} reads
\begin{equation}
\langle\cO_{\rm int}\cO_{\rm ext}\delta^a_{\rm\scriptscriptstyle
A}S\rangle = \langle \cO_{\rm ext}\delta^a_{\rm\scriptscriptstyle
A}\cO_{\rm int} \rangle\ .
\end{equation}
We now concentrate on the isovector component $a=1$. In our specific
application we choose $\cO_{\rm int}(x) = V_0^{1\rh}(x)$ for $x\in
R$ and $\cO_{\rm ext}(y) = P^{\rh 2}(y)$ for $y\notin R$, thus
obtaining
\begin{equation}
\label{naiveWI} \langle A_0^{2\rh}(x)P^{\rh 2}(y) \rangle = 2
\langle V_0^{1\rh}(x)P^{\rh 2}(y)\int_R {\rm d}^4z
\left\{\partial_\mu A_\mu^1 - 2mP^1\right\}\rangle \ .
\end{equation}
If we further require $R$ to be a time-oriented cylinder with periodic b.c. in space, i.e.
\begin{equation}
R = \{x:\ t_{1} \le x_0 \le t_2\},
\end{equation}
we immediately see that the space derivatives of the light axial
current on the right hand side of \Eq{naiveWI} drop out, while the
temporal derivative gives rise to a boundary contribution. After a
space integration of both sides over $\bx$, we arrive at our final
expression
\begin{equation}
\label{contWI}
 \langle Q_\rA^{2\rh}(x_0)P^{\rh 2}(y) \rangle =
2\langle Q_\rV^{1\rh}(x_0)P^{\rh 2}(y) \left\{\left[Q_\rA^1(t_2) -
Q_\rA^1(t_1)\right] - 2m\int_R {\rm d}^4z P^1(z)\right\}\rangle\ ,
\end{equation}
where $x_0\in[t_1,t_2]$, $y_0\notin[t_1,t_2]$ and we have introduced
the axial and vector charges
\begin{align}
Q_\rA^{a}(x_0) & = \int {\rm d}^3\bx\ A_0^{a}(x)\ ,  \\
Q_\rA^{k\rm h}(x_0) & = \int {\rm d}^3\bx\ A_0^{k\rh}(x)\ ,  \\
Q_\rV^{k\rm h}(x_0) & = \int {\rm d}^3\bx\ V_0^{k\rh}(x)\ .
\end{align}
\Eq{contWI} has to be understood as a relation among renormalized
quantities. It should be observed that $Q_\rA^{1}$ and $P^1$ consist
of two contributions in the flavour space, corresponding to the
non-zero matrix elements of $\tau^1$. Out of them, only those with
flavour content $\bar\psi_2\psi_1$ contribute to the right hand side
of the WI. These will be denoted respectively $Q_\rA^{21}$
and~$P^{21}$.


\section{Lattice implementation in the SF}

\begin{figure}[t]
  \begin{center}
    \epsfig{figure=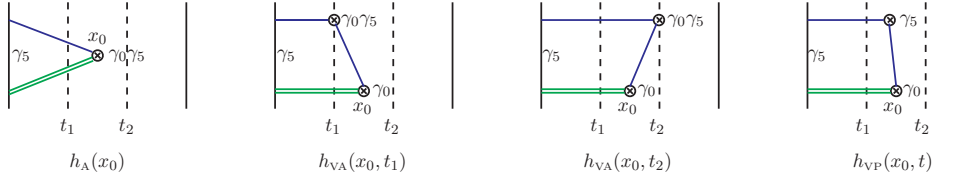, width=13.0 true cm}
  \end{center}
  \vskip -0.5cm
  \caption{Diagrammatic representation of the SF correlation functions of \protect\Eq{corrfuncs}. A single (double) line describes the propagation of a light (static) quark.  \label{fig:feynmantree}}
\end{figure}

The axial WI admits a straightforward lattice implementation. We
adopt here a SF topology where periodic boundary conditions (up
to a phase $\theta$ for the light-quark fields) are set up on the 
spatial directions and Dirichlet boundary conditions are
imposed on time at $x_0=0,T$. For a discussion of the original
application of the SF to the simplest WI, namely the PCAC, we refer
the reader to \cite{Luscher:1996sc}. Unexplained notation closely
follows \cite{Luscher:1996vw}.
\parbreak
Although the SF is formally defined in the continuum, we find it
convenient to work at finite lattice spacing. Light quarks are
assumed to be described by the ${\rm O}(a)$-improved Wilson action,
with the usual Sheikholeslami-Wohlert term in the bulk and
boundary counter-terms proportional to the improvement coefficients
$c_{\rm t}-1$ and $\tilde c_{\rm t}-1$. No background field is assumed. 
The static quark is instead defined in terms of the action
\begin{equation}
\label{staticaction} S_{\rm W}^{\rm stat}[\psi_{h},\bar
\psi_{h},\psi_{\bar h},\bar\psi_{\bar h},U] =
a^4\sum_x\left[\bar\psi_h(x)D_0^{{\rm W}*}\psi_h(x)-\bar\psi_{\bar
h} (x)D_0^{{\rm W}}\psi_{\bar h}(x)\right]\ ,
\end{equation}
where the forward and backward covariant derivatives
\begin{align}
D_0^{\rm W}\psi(x) & = \frac{1}{a}\left[W(x,0)
\psi(x+a\hat 0) - \psi(x)\right]\nonumber\ ,\\[2.0ex]
D_0^{\rm W*}\psi(x) & = \frac{1}{a}\left[\psi(x) -
W^\dagger(x-a\hat 0,0)\psi(x-a\hat 0)\right] \ ,
\end{align}
depend upon a parallel transporter $W$, which can be variously
defined. In this paper we consider four possible versions, namely
EH, APE, HYP1 and HYP2, respectively corresponding to
\begin{align}
\label{paralleltransp}
W^{\rm EH}(x,0)   & = U(x,0)\ , \nonumber \\[2.0ex]
W^{\rm APE}(x,0)  & = V(x,0)\ , \nonumber \\[2.0ex]
W^{\rm HYP1}(x,0) & = V^{\rm HYP}_{\vec\alpha}(x,0)\bigr|_{{\vec \alpha} = (0.75,0.6,0.3)} \ ,\nonumber \\[2.0ex]
W^{\rm HYP2}(x,0) & = V^{\rm HYP}_{\vec\alpha}(x,0)\bigr|_{{\vec \alpha} = (1.0,1.0,0.5)}\ .
\end{align}
In the above definitions $V(x,0)$ represents the average of the six
staples surrounding the gauge link $U(x,0)$, while $V^{\rm
HYP}(x,0)$ denotes the temporal HYP link of
\cite{Hasenfratz:2001hp}, with the approximate $SU(3)$ projection of
\cite{DellaMorte:2005yc}.
\parbreak
In order to translate \Eq{contWI} into the language of the SF, we
insert the static vector current in the middle of the bulk, i.e. at
$x_0=T/2$. The support region $R$ is then defined by localizing
$t_1$ and $t_2$ at different points, with the understanding that $0
< t_1 < x_0 < t_2 < T$ in order to avoid possible contact terms. The
pseudoscalar density is replaced by a boundary source uniformly
distributed over the spatial coordinates, i.e.
\begin{equation}
\label{hlsource} \Sigma^{\rh 2} = \frac{a^6}{L^3}\sum_{\by\bz}
\bar\zeta_\rh(\by)\gamma_5\zeta_2(\bz)\ .
\end{equation}
\parbreak
On-shell ${\rm O}(a)$-improvement of the quark currents requires the
introduction of operator counter-terms, whose structure has been
discussed in \cite{Luscher:1996sc,Kurth:2000ki}. Accordingly, we
introduce the ${\rm O}(a)$-improved currents
\begin{alignat}{4}
A_0^{ij;\rI}(x) & =  A_0^{ij}(x) + a\cA\delta A_0^{ij}(x)\ , & \qquad \delta A_0^{ij}(x) & =  \frac{1}{2}(\partial_0 + \partial_0^*)\bar\psi_i(x)\gamma_5\psi_j(x)\ ; \\[1.7ex]
A_0^{k\rh;\rI}(x) & =  A_0^{k\rh}(x) + a\cAstat\delta A_0^{k\rh}(x)\ , & \qquad \delta A_0^{k\rh}(x) & =  \bar\psi_k(x)\gamma_j\gamma_5\frac{1}{2}(\overleftarrow{\nabla}_j + \overleftarrow{\nabla}_j^*)\psi_\rh(x)\ ; \\[1.7ex]
V_0^{k\rh;\rI}(x) & =  V_0^{k\rh}(x) + a\cVstat\delta V_0^{k\rh}(x)\
, & \qquad \delta V_0^{k\rh}(x) & =
\bar\psi_k(x)\gamma_j\frac{1}{2}(\overleftarrow{\nabla}_j +
\overleftarrow{\nabla}_j^*)\psi_\rh(x)\ ;
\end{alignat}
and the ${\rm O}(a)$-improved charges
\begin{alignat}{3}
Q_\rA^{ij;\rI}(x_0) & = a^3\sum_\bx A_0^{ij;\rI}(x)  \ & = & \ Q_\rA^{ij}(x_0) + a\cA\delta Q_\rA^{ij}(x_0)\ ,  \\[1.1ex]
Q_\rA^{k\rh;\rI}(x_0) & = a^3\sum_\bx A_0^{k\rh;\rI}(x) \ & = & \ Q_\rA^{k\rh}(x_0) + a\cAstat\delta Q_\rA^{k\rh}(x_0)\ , \\[1.1ex]
Q_\rV^{k\rh;\rI}(x_0) & = a^3\sum_\bx V_0^{k\rh;\rI}(x) \ & = & \
Q_\rV^{k\rh}(x_0) + a\cVstat\delta Q_\rV^{k\rh}(x_0)\ ,
\end{alignat}
where, as also explained at the end of last section, the notation
$\cO^{ij}$ always refers to a flavour content $\bar\psi_i\psi_j$.
The improvement coefficients $\cA$, $\cAstat$ and $\cVstat$ depend
on the gauge coupling and are perturbatively expanded according to
\begin{equation}
c = c^{(1)}g_0^2 + c^{(2)}g_0^4 + {\rm O}(g_0^6)\ .
\end{equation}
\parbreak
In view of phenomenological applications, it is useful to allow for
renormalized currents at non-zero light-quark mass. ${\rm
O}(a)$-improvement requires in this case the introduction of
additional mass counter-terms, proportional to $\mq = m - \mcr$. The
relations between renormalized currents and their bare counterparts
explicitly read
\begin{align}
A_{0,\rm\scriptscriptstyle R}^{ij;\rI}(x) & = \ZA[1+\bA am_\rqq]A_0^{ij;\rI}(x)\ , \nonumber \\[1.7ex]
A_{0,\rm\scriptscriptstyle R}^{k\rh;\rI}(x) & = \ZAstat[1+\bAstat am_\rqq]A_0^{k\rh;\rI}(x)\ , \nonumber \\[1.7ex]
V_{0,\rm\scriptscriptstyle R}^{k\rh;\rI}(x) & = \ZVstat[1+\bVstat am_\rqq]V_0^{k\rh;\rI}(x)\ .
\end{align}
\parbreak
The SF implementation of the axial WI is then realized through the
introduction of a set of two- and three-point correlation functions,
\begin{align}
\label{corrfuncs}
h^{\rI}_{\rA}(x_0) & = \displaystyle{\langle Q^{2\rm{h};\rI}_\rA(x_0)\Sigma^{\rm{h}2}\rangle}\ , \nonumber \\[1.7ex]
h^{\rI}_{\rVA}(x_0,y_0) & = \displaystyle{\langle Q_\rV^{1\rm{h};\rI}(x_0)Q^{21;\rI}_\rA(y_0)\Sigma^{\rm{h}2}\rangle}\ , \nonumber \\[1.3ex]
h^{\rI}_{\rVP}(x_0,y_0) & = \frac{a^3}{L^3}\,\sum_{\by }
\displaystyle{\langle
Q^{1\rm{h};\rI}_\rV(x_0)P^{21}(y)\Sigma^{\rm{h}2}\rangle} \ ,
\end{align}
which are graphically represented by the Feynman diagrams of
Fig.~\ref{fig:feynmantree}. It should be observed that the two-point
correlator $h^{\rI}_{\rA}$ satisfies the relation
$h^{\rI}_{\rA}=-2f_\rA^{\rm{stat,I}}$ with $f_\rA^{\rm{stat,I}}$
defined in Eqs.~(3.22-3.24) of \cite{Kurth:2000ki}. Once the
renormalized currents are expressed in terms of the bare ones, the
axial WI takes the form of a constraining relation among
renormalization constants. In the chiral limit it reduces to
\begin{equation}
  \label{sfwardid}
 \cR\equiv \frac{h^\rI_{\rVA}(x_0,t_2) - h^\rI_{\rVA}(x_0,t_1)}{h^\rI_\rA(x_0)} = \frac{\ZAstat}{\ZVstat\ZA} + {\rm O}(a^2)   \ .
\end{equation}
\parbreak
In order to pursue a numerical implementation of \Eq{sfwardid}, some
geometrical parameters have to be specified, namely the ratios
$T/L$, $t_1/T$, $t_2/T$ and the $\theta$-angle of the SF. Concerning
the latter, we  consider three possible values, i.e. $\theta = 0.0,
0.5, 1.0$. The other parameters will instead be collectively
referred to as the topology $\cT$ of the WI. In
Table~\ref{tab:topologies} we list four possibilities. Each of them
affects the noise-to-signal ratio of the non-perturbative
simulations in its own way and introduces specific cutoff effects in
the ratio $\cR$ at finite lattice spacing. Therefore, a convenient
choice of $\cT$ imposes -- at least theoretically~-- a balance
between the minimization of the lattice artefacts and the
maximization of the numerical signal.
\begin{table}[!ht]
    \begin{center}
      \begin{tabular}{ccccc}
        \\\hline\hline\\[-2.0ex]
        $\cal T$ & $T/L$ & $x_0/T$ & $t_1/T$ & $t_2/T$ \\[0.5ex]
        \hline\\[-2.0ex]
        $1$ & $1$   & $1/2$ & $1/4$ & $3/4$ \\
        $2$ & $2$   & $1/2$ & $1/4$ & $3/4$ \\
        $3$ & $3/2$ & $1/2$ & $1/3$ & $2/3$ \\
        $4$ & $3$   & $1/2$ & $1/3$ & $2/3$ \\[1.0ex]
        \hline\hline
      \end{tabular}
      \vskip 0.1cm
  \caption{Some topologies $\cal T$ of the WI.\label{tab:topologies}}
  \end{center}
\end{table}
\parbreak
We remark that \Eq{sfwardid}, which we use in order to determine
$\cVstat$, depends as well on the improvement coefficients $\cA$ and
$\cAstat$. These have been determined respectively in
\cite{Luscher:1996ug} and \cite{DellaMorte:2005yc} and are taken as
input parameters here. In particular, $\cAstat$ is analytically
known at one-loop order for the EH and APE actions, and effectively
up to ${\rm O}(g_0^4)$-terms for the HYP1 and HYP2 actions. Scaling
tests of $\cAstat$ have been extensively discussed in
\cite{DellaMorte:2005yc}, to which the reader is referred for
details. Here we stress that the lack of a full knowledge of
$\cAstat$ introduces systematic uncertainties at order ${\rm
O}(g_0^4)$ in the determination of $\cVstat$. On the other hand, 
the WI is independent of the boundary improvement coefficients $\ct$ and 
$\cttilde$. This has been explicitely checked in perturbation theory.


\section{One-loop perturbative analysis of the WI}

\begin{figure}[t]
  \label{oneloopW}
  \begin{center}
    \epsfig{figure=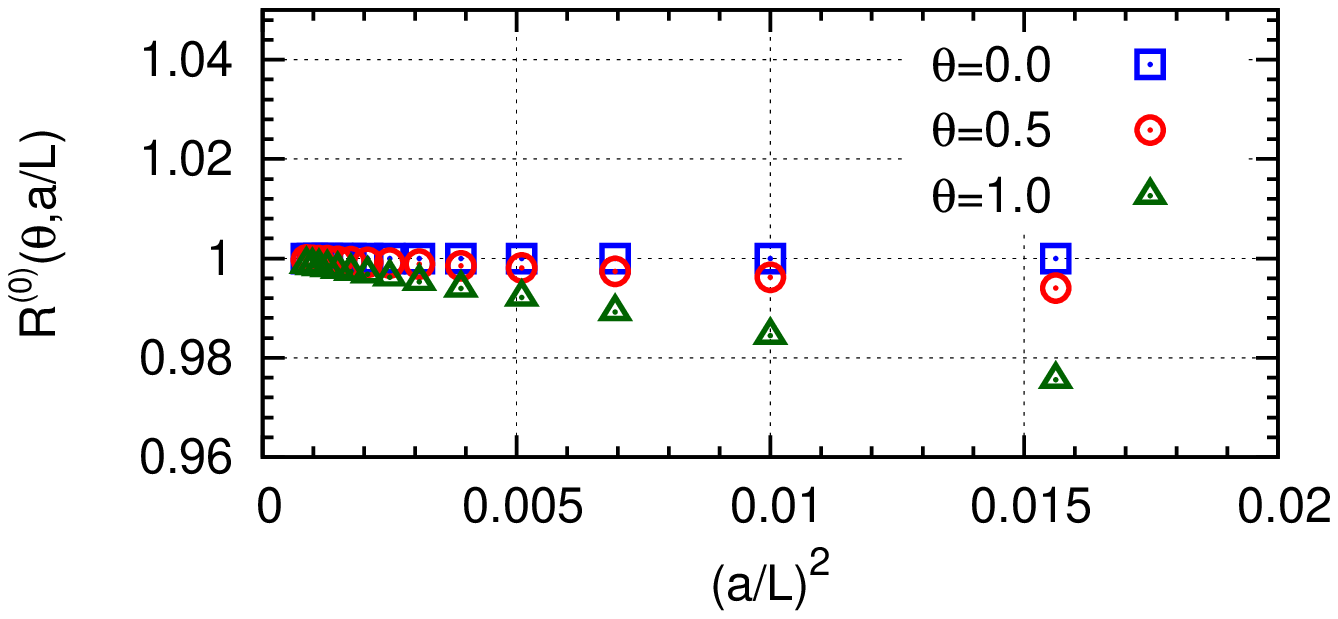, angle=-90, width=8.0 true cm}
  \end{center}
  \vskip -0.6cm
  \caption{Continuum approach of $\cR^{(0)}$ with topology $\cT=2$.}
  \label{fig:treecont}
\end{figure}

\begin{figure}[t]
  \hskip 0.2cm
  \epsfig{figure=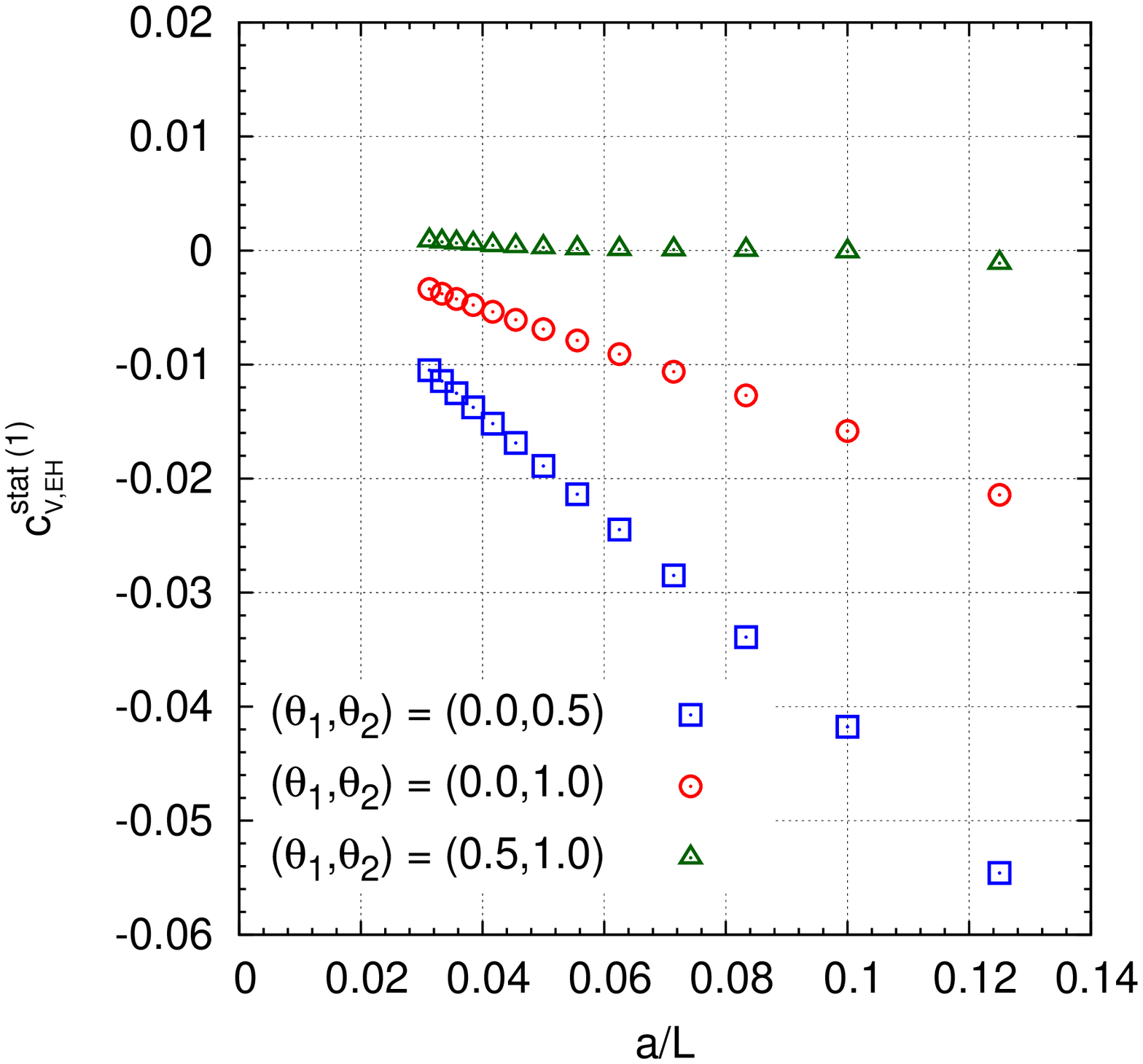, angle=-90, width=7 true cm}
  \hskip 0.5cm
  \epsfig{figure=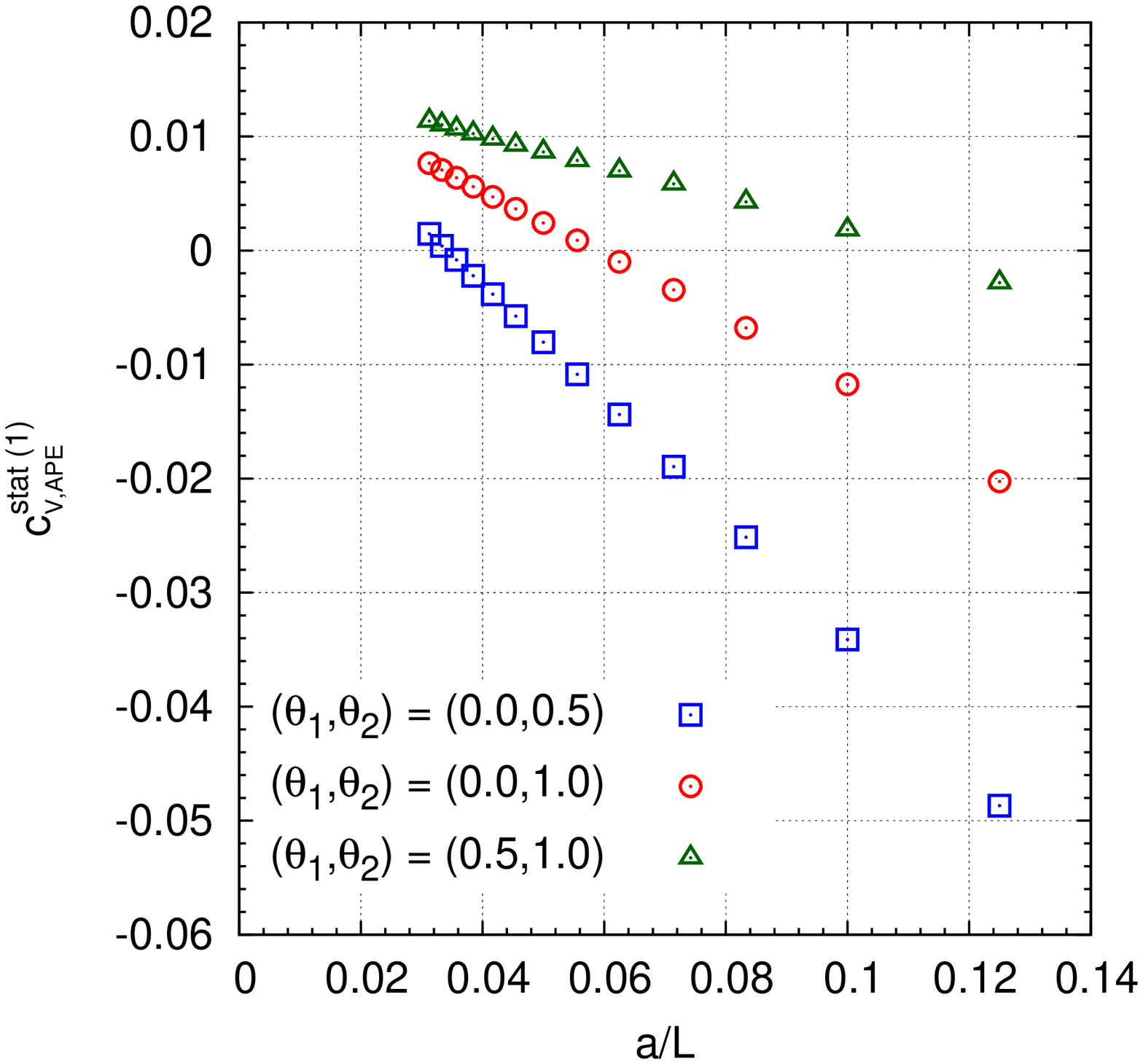, angle=-90, width=7 true cm}
  \caption{Continuum approach of $\cVstatone$ for the EH (left plot)
    and APE (right plot) actions. Plots refer to the topology $\cT=2$.
    Different choices of the $\theta$-angles provide independent
    definitions of $\cVstatone$. Plotted points correspond to
    $L/a=8,\dots,32$.\label{fig:cvstat1}}
\end{figure}

\begin{figure}[h]
  \begin{center}
    \hskip 0.2cm \hbox{\hskip
      -0.2cm}\epsfig{figure=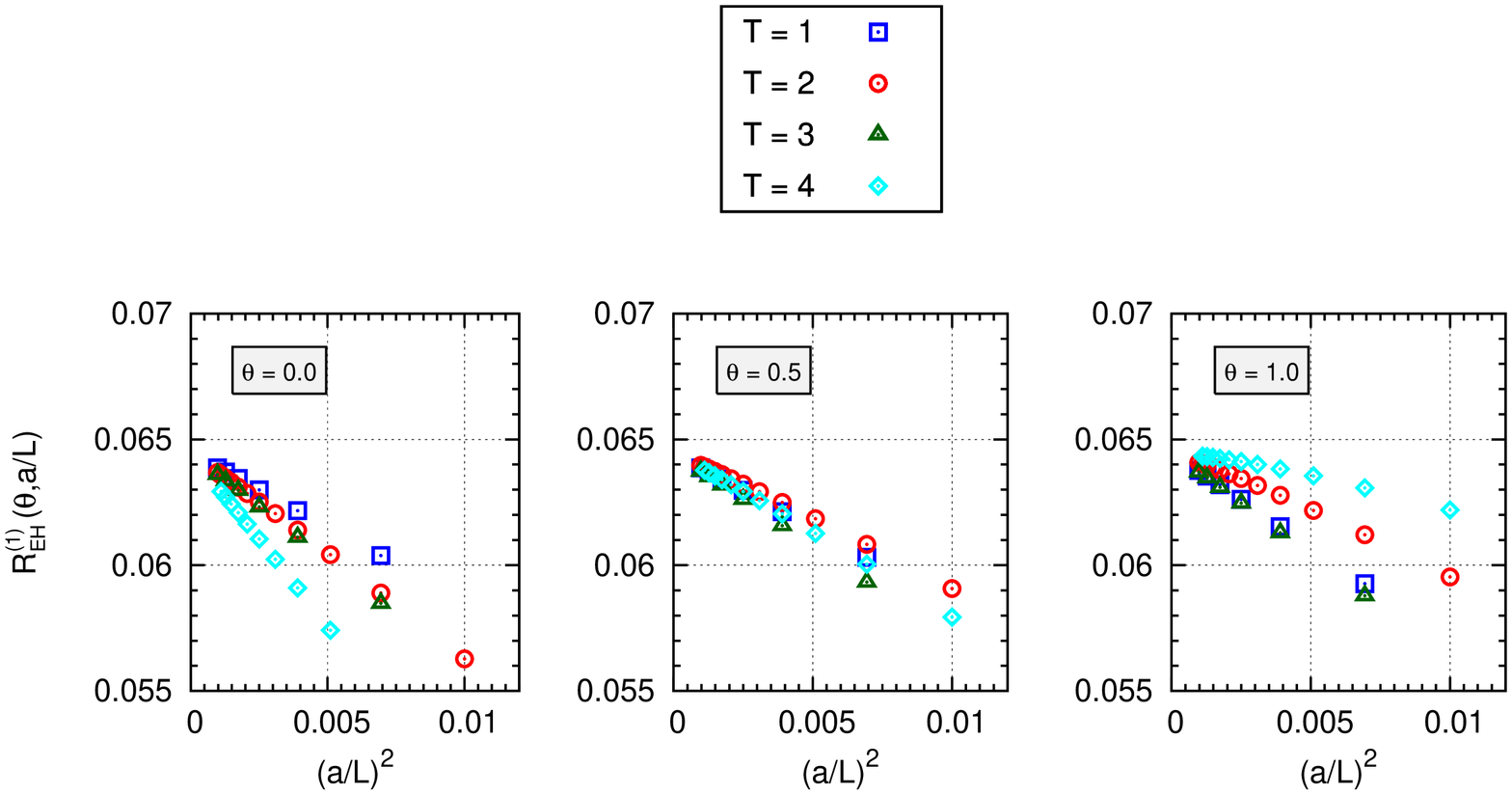, angle=-90,
      width=14.5 true cm}
    \\
    \hbox{\hskip 2.2 true cm}
    \epsfig{figure=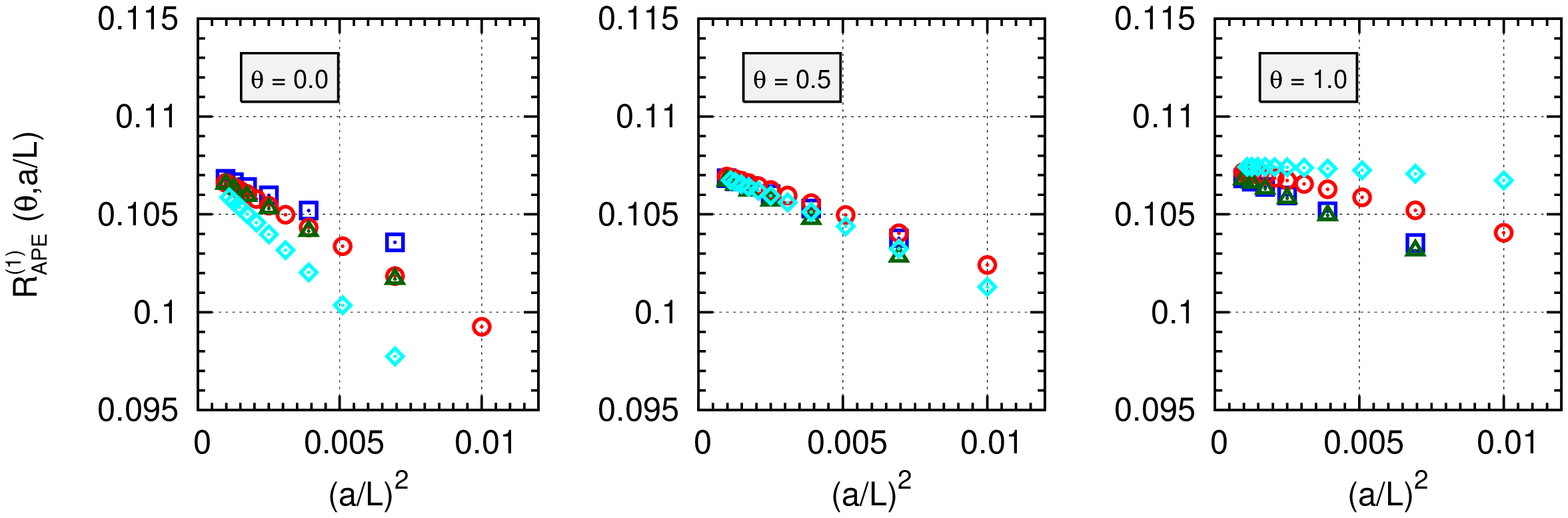, angle=-90, width=14.5
      true cm}
  \end{center}
  \caption{Continuum approach of $\cR^{(1)}$ for the EH (upper plots)
    and APE (lower plots) actions. Plots refer to various SF topologies
    and $\theta$ angles. Plotted points correspond to $L/a=10,\dots,32$
    for $\cT=2,4$ and $L/a=12,\dots,32$ for $\cT=1,3$.\label{fig:R1impr}}
\end{figure}

A first indication of the cutoff effects related to a given choice
of the topology $\cT$ can be obtained in principle from a one-loop
perturbative calculation of the WI. We anticipate that once the
${\rm O}(a)$-improvement has been carried out, the residual lattice
artefacts of ${\rm O}(a^2)$ have comparable size in the various
topologies, so that a conclusive argument for the choice of the
preferred $\cT$ has to follow from non-perturbative
considerations. To show this, we expand the ratio $\cR$ in powers of
the coupling, i.e.\begin{equation} \label{Rexpanded} \cR = \cR^{(0)}
  + g_0^2\cR^{(1)} + {\rm O}(g_0^4)\ .
\end{equation}
Each term of the perturbative expansion is a function of the bare
quark mass $m$ and must be computed at $m=\mcr$. Since the latter
depends in turn upon the bare coupling, each correlator $h$ of
\Eq{corrfuncs} has to be expanded according to
\begin{equation}
  \label{expcorr} h = h^{(0)}|_{m=0} + g_0^2\left[h^{(1)} + m_{\rm
      cr}^{(1)}\partial_m h^{(0)} + h_{\rm\scriptscriptstyle
      b}^{(1)}\right]_{m=0} + {\rm O}(g_0^4)\ ,
\end{equation}
where $\partial_m$ indicates a partial derivative with respect to $m$
and the subscript ``b'' denotes the contribution of the boundary
counter-terms proportional to $\tilde c_{\rm t}-1$. The one-loop
critical mass $m_{\rm cr}^{(1)}$ is defined here by requesting that
the ${\rm O}(a)$-improved PCAC quark mass vanish. Its values at
finite lattice spacing are taken from
\cite{Palombi:2005zd,Palombi:2006pu}.
\parbreak
The ratio $\cR$ is expected to be tree-level improved, since all the
improvement counter-terms start at ${\rm O}(g_0^2)$. This
expectation is confirmed by Fig.~\ref{fig:treecont}, where the
approach of $\cR^{(0)}$ to the continuum limit is reported for the
topology $\cT=2$. We observe that the slope of $\cR^{(0)}$ increases
with $\theta$ (the scaling is perfect at $\theta=0.0$) and is
independent of $\cT$. It follows that, in order to identify a better
$\cT$, at least the one-loop contribution has to be worked out
explicitly.
\parbreak
The perturbative expansion of the two-point correlator $h^{\rm
  I}_\rA$ has been discussed in \cite{Kurth:2000ki} and will not be
reviewed here. The one-loop coefficient of the three-point
correlator $h^{{\rm I}(1)}_{\rVA}$ receives contributions from
Feynman diagrams corresponding to self-energy and tadpole
corrections of the single quark legs, plus vertex corrections with
gluons propagating from one leg to another.
\parbreak
Several possible improvement conditions may be imposed in order to
tune $\cVstatone$ so that the ${\rm O}(a)$-improvement is realized
at one-loop order. After some attempts, we found that a reasonable
definition is to enforce the equation
\begin{equation}
  \label{Rimprcond}
  \cR^{(1)}(\theta_1,a/L) = \cR^{(1)}(\theta_2,a/L) + {\rm O}\left[(a/L)^2\right]\ ,
\end{equation}
which defines $\cVstatone$ up to ${\rm O}(a/L)$-terms. Cutoff
effects with the WI topology $\cT=2$ are reported in
Tables~\ref{tab:cvstatEHone}-\ref{tab:cvstatAPEone} and in
Fig.~\ref{fig:cvstat1} for the EH (left plot) and APE (right plot)
actions and three possible choices of the pair
$(\theta_1,\theta_2)$. \TABLE[r]{
  \small
  \centering
  \vbox{\vskip 1.2cm}
  \begin{tabular}{|c|c|c|}
    \hline
    & EH & APE \\
    \hline
    $\cVstatone$ & 0.0048(3) & 0.0185(3) \\
    \hline
  \end{tabular}
  \caption{$\cVstatone$ for the EH and APE actions.\label{tab:cvstatone}}
}
\noindent The other topologies show similar lattice artefacts. As
expected, different definitions converge to the same continuum
limit, which is very small for the EH discretization, if compared to
the size of the cutoff effects, and somewhat larger for the APE
action. It follows that the extrapolation of the lattice points to
the continuum is difficult and one should not expect a high
numerical precision. In order to reduce the size of the lattice
artefacts, we have employed the blocking
procedures of \cite{Luscher:1985wf,Palombi:2002gw}. Results are
reported in Table~\ref{tab:cvstatone}. Our determination of $\cVstatoneEH$
is in good agreement with the original estimate given by
\cite{Morningstar:1998yx} in the framework of NRQCD.
\parbreak
Once the improvement coefficient $\cVstatone$ is known, the ratio
$\cR^{(1)}$ can be calculated in the ${\rm O}(a)$-improved theory.
In Fig.~\ref{fig:R1impr} its continuum approach is plotted vs.
$(a/L)^2$ for all the topologies and the $\theta$-angles.
Corresponding data are reported in Tables
\ref{tab:REHone}-\ref{tab:RAPEone}. The main feature of the plots is
the similarity of the various definitions, which differ by just a
few percent at the coarsest lattices. Nevertheless, some topologies
are more sensitive to a change of $\theta$ than others, e.g. $\cT=4$
looks almost flat at $\theta=1.0$, while it has the largest slope at
$\theta=0.0$. Remarkably, the
\TABLE[r]{
  \small
  \centering
  \vbox{\vskip -0.6cm}
  \begin{tabular}{|c|c|c|}
    \hline
    & EH & APE \\
    \hline
    $\ZAstat/\ZVstat$ & $1-0.0521(1)g_0^2$ & $1-0.0093(2)g_0^2$ \\
    \hline
  \end{tabular}
  \caption{$\ZAstat/\ZVstat$ at one-loop order for the EH and APE actions.\label{tab:ZAZV}}
}
\noindent spread between different $\cT$'s almost vanishes around
$\theta=0.5$, thus suggesting that this $\theta$-value could be the
most stable against variations of the topology beyond perturbation
theory.
\parbreak
We extract the common continuum limit of $\cR^{(1)}$ via the
afore-mentioned blocking techniques. Our best estimates are
$\cR^{(1)}_{\scriptscriptstyle\rm EH}~=~0.0644(1)$ and $
\cR^{(1)}_{\scriptscriptstyle\rm APE} = 0.1072(2)$. In order to
isolate the ratio of the static renormalization constants
$\ZAstat/\ZVstat$, the one-loop contribution of $\ZA$, i.e.
$\ZA^{(1)} = -0.116458$ \cite{Gabrielli:1990us,Sint:private}, has to
be subtracted from $\cR^{(1)}$. Results are reported in
Table~\ref{tab:ZAZV}. The value obtained with the EH action is not
novel: it checks the one previously found in \cite{Borrelli:1992fy,Hashimoto:2002pe}
within 3\%.


\section{Non-perturbative determination of $\ZAstat/\ZVstat$}

\begin{figure}[t]
  \begin{center}
    \hskip -0.5cm
    \epsfig{figure=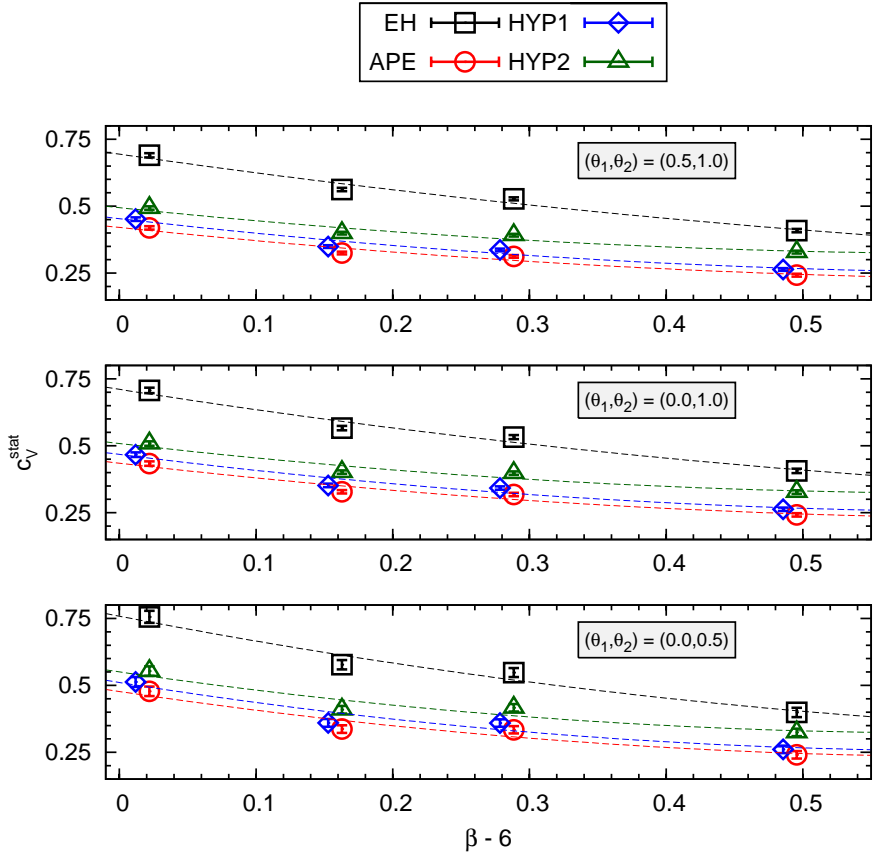, angle=-90, width=11.5 true cm}
  \end{center}
  \vskip -0.4cm
  \caption{
    $\beta$-dependence of $\cVstat$ for some choices of the pair 
    $(\theta_1,\theta_2)$ and of the static action. 
    For the sake of readability, points (diamonds) have been slightly shifted along 
    the horizontal axis. Dashed curves represent quadratic fits. 
    \label{fig:cvstatnp}}
  \vskip -0.3cm
\end{figure}

\begin{figure}[t]
\begin{center}
\hskip -1.0cm\epsfig{figure=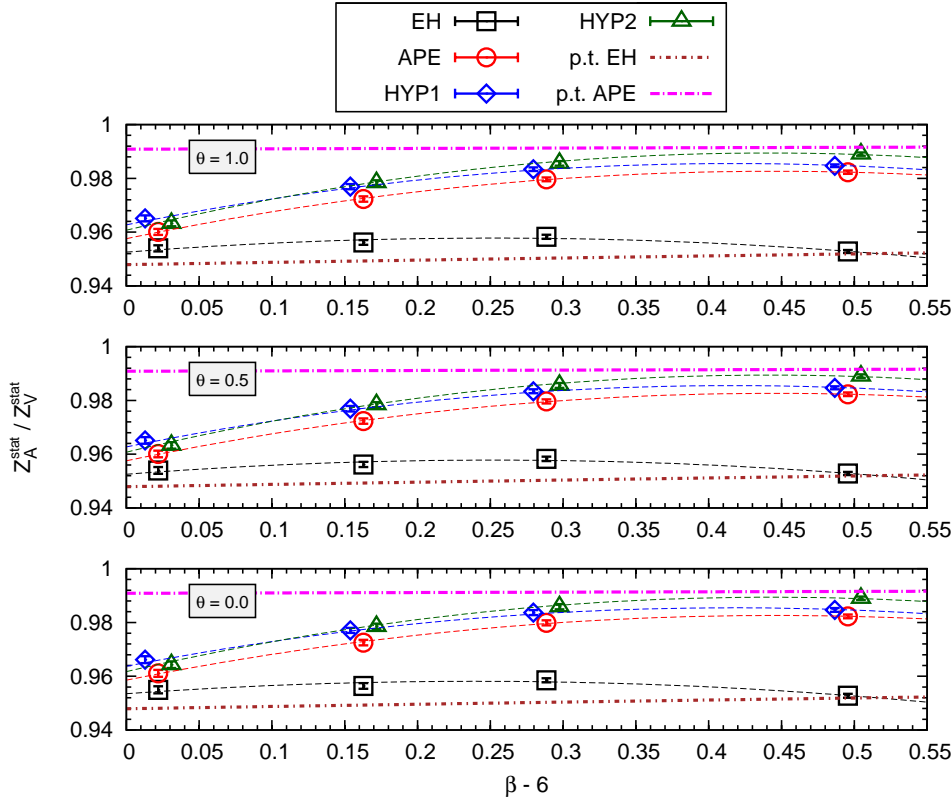, angle=-90, width=12.5 true cm}
\end{center}
\vskip -0.4cm
\caption{${\rm O}(a)$-improved ratio of the renormalization
  constants $\ZAstat/\ZVstat$ for various static
  actions. Different choices of $\theta$ correspond to independent definitions of the WI. 
  For the sake of readability, points (diamonds and triangles) have been slightly 
  shifted along the horizontal axis. Dashed curves represent quadratic fits.
  \label{fig:zazvimprnp}}
\end{figure}

In order to simulate the WI non-perturbatively, we first address the
choice of the geometrical parameters. Some numerical attempts
suggest that the topology with the best signal-to-noise ratio is
also the one with the smallest aspect ratio $T/L$. This is largely 
expected on the basis of \cite{DellaMorte:2005yc}, since the loss 
of signal is mainly related to the temporal extension of the static
propagator, which for every $\cT$ goes from the boundary to
$x_0=T/2$. Given the lack of a clear indication from perturbation
theory concerning the preeminence of a specific topology over the
others, we decide to just follow the criterion of the
signal-to-noise ratio and to consequently adopt $\cT=1$ for our
non-perturbative study. Simulation parameters are collected in
Table~\ref{tab:simpar}. They have been taken from
\cite{DellaMorte:2005yc} and correspond to a physical size
$L=2L_{\rm max} = 1.436r_0$ of the SF.
\begin{table}[ht]
    \begin{center}
      \begin{tabular}{cccc}
        \\\hline\hline\\[-2.0ex]
        $L/a$ & $\beta$ & $\kappa$ & $\theta$  \\ \\[-2.5ex]
        \hline\\[-2.0ex]
         8 & 6.0219 & 0.135081,0.1344011 & 0.0,0.5,1.0 \\
        10 & 6.1628 & 0.135647,0.1351239 & 0.0,0.5,1.0 \\
        12 & 6.2885 & 0.135750,0.1353237 & 0.0,0.5,1.0 \\
        16 & 6.4956 & 0.135593,0.1352809 & 0.0,0.5,1.0 \\[0.5ex]
        \hline\hline
      \end{tabular}
      \vskip 0.1cm
  \caption{Simulation parameters used for the non-perturbative study of the
    improvement coefficients $\cVstat$, $\bVstat$ and the ratio
    $\ZAstat/\ZVstat$.\label{tab:simpar}}
  \end{center}
  \vskip -0.5cm
\end{table}
Moreover, $\csw$ is non-perturbatively tuned according to
\cite{Luscher:1996ug} and the boundary improvement coefficients
$\ct$ and $\cttilde$ are respectively set to their two- and 
one-loop values \cite{Luscher:1996vw}. 
\parbreak
It is worth noting that with $\cT=1$ the WI can be
simulated directly at the chiral point, with no need for a mass
extrapolation in the way of \cite{DellaMorte:2005rd}. The
$\kappa$-values which have been used are the ones reported on the
left of the third column and correspond to $\kappa_{\rm cr}$
obtained from the ${\rm O(a)}$-improved PCAC relation
\cite{Heitger:2003xg,Capitani:1998mq}. 
\parbreak
We also observe that the simulation at $\beta=6.1628$ cannot be
actually performed with $t_1=T/4$ and $t_2=3T/4$, since these are
non-integer multiples of the lattice spacing in this particular
case. To avoid the problem, we take here $t_1=2a$ and $t_2=7a$. This
choice is theoretically sound since no contact term turns up in the
WI. It amounts to changing the definition of $\cVstat$ by an ${\rm
O}(a)$-term and of the improved ratio $\ZAstat/\ZVstat$ by an ${\rm
O}(a^2)$-term at that given $\beta$. Other choices are possible as
well. The present one has the {\it a posteriori} advantage that it
makes the  $\beta$-dependence of $\ZAstat/\ZVstat$ smoother than
other definitions.
\parbreak
To achieve a non-perturbative estimate of $\ZAstat/\ZVstat$ we first
have to properly tune the improvement coefficient $\cVstat$. We
follow the perturbative definition introduced in \Eq{Rimprcond} and
impose the improvement condition
\begin{equation}
\label{imprcondnp}
\cR(\theta_1,\beta) = \cR(\theta_2,\beta) + {\rm O}(a^2)\ ,
\end{equation}
where, as already explained in sect.~3, the coefficients $\cAstat$
and $\cA$ are taken as input parameters. We stress once more that since 
$\cAstat$ is known up to ${\rm O}(g_0^4)$-terms, this introduces a systematic 
uncertainty in our computation, thus making the numerical estimate of 
$\cVstat$ only non-perturbatively effective. In principle, it could be
possible to avoid this by enforcing two simultaneous conditions,
i.e.
\begin{align}
\cR(\theta_1,\beta) & = \cR(\theta_2,\beta) + {\rm O}(a^2)\ , \nonumber \\
\cR(\theta_1,\beta) & = \cR(\theta_3,\beta) + {\rm O}(a^2)\ , \qquad \theta_1 \ne \theta_2 \ne \theta_3 \ ,
\end{align}
from which $\cAstat$ and $\cVstat$ could be determined at the same
time with no approximation. Unfortunately, the resulting expressions
for the improvement coefficients are quite involved and
characterized by a very poor signal. For this reason we are forced
to resort to \Eq{imprcondnp}. Being $\cR$ linearly dependent on
$\cVstat$, i.e.
\begin{equation}
\cR(\theta,\beta) = {r}(\theta,\beta) + \cVstat(\beta) {s}(\theta,
\beta)\ ,
\end{equation}
we obtain the improvement coefficient from the equation
\begin{equation}
\cVstat(\beta) = - \frac{ {r}(\theta_1,\beta) - {r}(\theta_2,\beta)
} { {s}(\theta_1,\beta) - {s}(\theta_2,\beta) }\ +\ {\rm O}(a)\ .
\end{equation}
Results at the simulation points are reported in
Table~\ref{tab:cvstatnp} of appendix A. Statistical errors have been
computed through the jackknife method. In Fig.~\ref{fig:cvstatnp}
the $\beta$-dependence of $\cVstat$ is shown for different
choices of the angles $(\theta_1,\theta_2)$ and for different static
discretizations. The most noticeable feature seems to be the large
discrepancy with respect to the perturbative estimates given in the
previous section. We also observe that the EH determination is quite
distinct from the other regularizations, which are instead close to
each other.
\vskip 0.05cm
The numerical values of $\cVstat$ should be independent of the
choice of $(\theta_1,\theta_2)$ up to ${\rm O}(a)$-effects. Therefore,
the difference $\Delta\cVstat = \cVstat|_{(\theta_1,\theta_2)}
-\cVstat|_{(\theta'_1,\theta'_2)}$ is expected to decrease at larger
$\beta$-values. This is confirmed by our data.
\vskip 0.05cm
As Table~\ref{tab:cvstatnp} shows, the improvement condition with
the best signal-to-noise ratio is the one corresponding to 
$(\theta_1,\theta_2)=(0.5,1.0)$. This is also the one with the smallest 
perturbative cutoff effects. A quadratic fit of it in the range of the 
Monte Carlo simulations ($6.0\le \beta \le 6.5$) leads to the parametrization
\begin{alignat}{3}
&\cVstatEH     & & \ = \ 0.694 - 0.732x + 0.330x^2\ , \label{eq:cvstatnp1} \\[1.0ex]
&\cVstatAPE    & & \ = \ 0.421 - 0.531x + 0.360x^2\ , 
\end{alignat}
\begin{alignat}{3}
&\cVstatHYPone & & \ = \ 0.453 - 0.584x + 0.421x^2\ ,   \\[1.0ex]
&\cVstatHYPtwo & & \ = \ 0.494 - 0.528x + 0.404x^2\ ;\qquad x = \beta-6 \ .  \label{eq:cvstatnp4}
\end{alignat}
\parbreak
The ${\rm O}(a)$-improved ratio of the renormalization constants
$\ZAstat/\ZVstat$ corresponding to this choice of $\cVstat$ is shown 
in Fig.~\ref{fig:zazvimprnp}; the same data are collected in 
Table~\ref{tab:zazvnp}. To extract this ratio out of $\cR$, we used 
the ALPHA determination of $\ZA$ reported in \cite{Luscher:1996jn}. 
Since now all the improvement counter-terms have been taken into account, 
the definition of $\ZAstat/\ZVstat$ with $\theta=0.0$ 
has to agree up to ${\rm O}(a^2)$-terms with those at $\theta=0.5$ and 
$\theta=1.0$, which have been used in order to tune the improvement coefficient. 
Indeed, it can be seen from Table~\ref{tab:zazvnp} that the differences 
are zero within the statistical errors. 
Aside the non-perturbative determination, also the one-loop estimates of 
Table~\ref{tab:ZAZV} are reported in Fig.~\ref{fig:zazvimprnp}. The agreement 
is good with EH static fermions in the whole region explored by the Monte 
Carlo simulations. It is good as well with the APE action at the largest
$\beta$-values. A quadratic fit (in the range $6.0\le \beta \le 6.5$)
gives
\begin{alignat}{3}
& \ZAZVEH(g_0)     & & \ = \  0.953 + 0.0417 x - 0.0828 x^2\ , \label{eq:zvstatnp1}  \\[1.0ex]
& \ZAZVAPE(g_0)    & & \ = \  0.958 + 0.113  x - 0.126  x^2\ ,  
\end{alignat}
\begin{alignat}{3}
& \ZAZVHYPone(g_0) & & \ = \  0.963 + 0.109  x - 0.131  x^2\ ,   \\[1.0ex]
& \ZAZVHYPtwo(g_0) & & \ = \  0.961 + 0.129  x - 0.146  x^2\ ;\qquad x = \beta-6 \ . \label{eq:zvstatnp4}
\end{alignat}
Eqs.~(\ref{eq:zvstatnp1}-\ref{eq:zvstatnp4}) reproduce the numbers
of Table~\ref{tab:zazvnp} within the statistical errors.  For
convenience, we also report a parametrization of $\ZAstat/\ZVstat$
with all the improvement coefficients set to their respective
values, but $\cAstat=\cVstat=0$. With this choice, the WI is not
${\rm O}(a)$-improved. Definitions corresponding to different
choices of the $\theta$-angle differ now by ${\rm O}(a)$-terms, which
are well above the statistical uncertainties. At $\theta = 0.5$ we
find
\begin{alignat}{3}
& \ZAZVEH(g_0)     & & \ = \  0.818 + 0.186 x - 0.134 x^2\ ,   \label{eq:zvstatnp1cVzero}  \\[1.0ex]
& \ZAZVAPE(g_0)    & & \ = \  0.917 + 0.176 x - 0.165 x^2\ ,  \\[1.0ex]
& \ZAZVHYPone(g_0) & & \ = \  0.926 + 0.178 x - 0.177 x^2\ ,  \\[1.0ex]
& \ZAZVHYPtwo(g_0) & & \ = \  0.978 + 0.158 x - 0.179 x^2\ ;  \qquad x = \beta-6 \ . \label{eq:zvstatnp4cVzero}
\end{alignat}

Our final results, represented by Eqs.~(\ref{eq:zvstatnp1})-(\ref{eq:zvstatnp4}) 
depend upon the choice of $c_{\scriptscriptstyle\rm A}^{\rm stat}$. Adopting
the determination of \cite{DellaMorte:2005yc} introduces a systematic uncertainty at
${\rm O}(g_0^4)$, which propagates to the ratios
$Z_{\rm\scriptscriptstyle A}^{\rm stat}/Z_{\rm\scriptscriptstyle V}^{\rm stat}$
and can be easily estimated. We first observe that the improvement
coefficient $c_{\rm\scriptscriptstyle V}^{\rm stat}$ is very sensitive
to variations of $c_{\rm\scriptscriptstyle A}^{\rm stat}$. This is not
surprising, since both the counter-terms proportional to
$c_{\rm\scriptscriptstyle V}^{\rm stat}$ and
$c_{\rm\scriptscriptstyle A}^{\rm stat}$ are meant to cancel the ${\rm O}(a)$-lattice
artefacts of the WI. Therefore, a change of ${\rm O}(1)$ in
$c_{\rm\scriptscriptstyle A}^{\rm stat}$ is expected to produce a variation of
the same order in $c_{\rm\scriptscriptstyle V}^{\rm stat}$ via Eq.~(\ref{imprcondnp}). 
In practice, setting $c_{\rm\scriptscriptstyle A}^{\rm stat}=0$ lowers the estimates of
Eqs.~(\ref{eq:cvstatnp1})-(\ref{eq:cvstatnp4}) by $30\%$ at the coarsest lattice 
spacing. Nevertheless, if the new values of 
$c_{\rm\scriptscriptstyle V}^{\rm stat}|_{c_{\rm\scriptscriptstyle A}^{\rm stat}=0}$
are introduced in the WI and the counter-term of the static axial current is
explicitly dropped out at denominator of Eq.~(\ref{sfwardid}), the variation of
$Z_{\rm\scriptscriptstyle A}^{\rm stat}/Z_{\rm\scriptscriptstyle V}^{\rm stat}$
amounts to at most $1\%$. This is due to a very large numerical cancellation of
$c_{\rm\scriptscriptstyle A}^{\rm stat}$ within the ratio ${\cal R}$.
It follows that Eqs.~(\ref{eq:zvstatnp1})-(\ref{eq:zvstatnp4}) can be assigned a systematic uncertainty
of $1\%$.
\parbreak
Concerning the systematic uncertainty of $\cVstat$, which has a strong dependence
upon $\cAstat$ in our determination, we naively expect that physical matrix elements 
of the static vector current will be only slightly affected by variations of $\cVstat$, 
in strict analogy with the static axial current, where a change of the operator 
counter-term is compensated by an opposite variation in the renormalization constant, 
as shown in \cite{DellaMorte:2007ij}. Unfortunately, we have no quantitative elements
at the moment to clarify if this is the case also for the static vector current. 


\section{The improvement coefficient $\bVstat$}

\FIGURE[!hr]{
  \small
  \centering
  \epsfig{figure=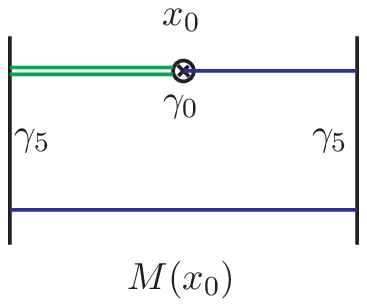, angle=0, width=2.5 true cm}
  \caption{Three-point SF correlator.\label{fig:threepoint}}
}
\noindent 
The axial WI at non-zero light-quark mass is complicated by the
presence of a mass term proportional to the temporal integral of the
SF correlator $h_{\rVP}^{\rm\scriptscriptstyle I}$ introduced in
\Eq{corrfuncs}. Since the integration region covers the whole
interval $[t_1,t_2]$, an integrable contact term raises at 
$y_0=x_0=T/2$. Managing the integral can be disadvantageous
in some cases: for instance, in perturbation theory it requires a
complete one-loop calculation for each value of the integration
variable, since no Fourier transform is defined in the SF along the
time direction. Therefore, in order to improve the static vector
current out of the chiral limit, it is easier to look for some more
comfortable observable.
\parbreak
One attractive possibility is to consider a three-point SF
correlator with the insertion of the static vector current in the
bulk. To this aim we define
\begin{equation}
M^\rI(x_0,m) = \langle {\Sigma'}^{21} V_0^{1\rm h;I}(x) \Sigma^{\rh2}\rangle \ ,
\end{equation}
where $\Sigma^{\rh2}$ has been introduced in \Eq{hlsource} and
\begin{equation}
{\Sigma'}^{21} = \frac{a^6}{L^3}\sum_{\by'\bz'} \bar\zeta'_2(\by')\gamma_5\zeta'_1(\bz')
\end{equation}
is a relativistic pseudoscalar boundary source localized
at $x_0=T$. Since we are interested in massive light-quarks, we keep
the mass dependence explicit in the definition of $M^\rI$. The flavour
structure of the chosen valence operators allows for just one Wick
contraction, depicted in Fig.~\ref{fig:threepoint}. To have it
renormalized, all the logarithmic divergences must be subtracted,
both those related to the static vector current and the ones induced
by the boundary sources. In the ${\rm O}(a)$-improved theory the
renormalized correlator reads
\vskip 0.2cm
\begin{equation}
M^\rI_{\scriptscriptstyle\rm R}(x_0,m_{\rm\scriptscriptstyle R}) = \ZVstat Z_\zeta^3Z_\zeta^{\rh}\{1+b_\zeta a\mq\}^3
\{1+\bVstat a\mq\}M^\rI(x_0,m)\ .
\end{equation}
\begin{figure}[t]
\begin{center}
\hskip -1.0cm\epsfig{figure=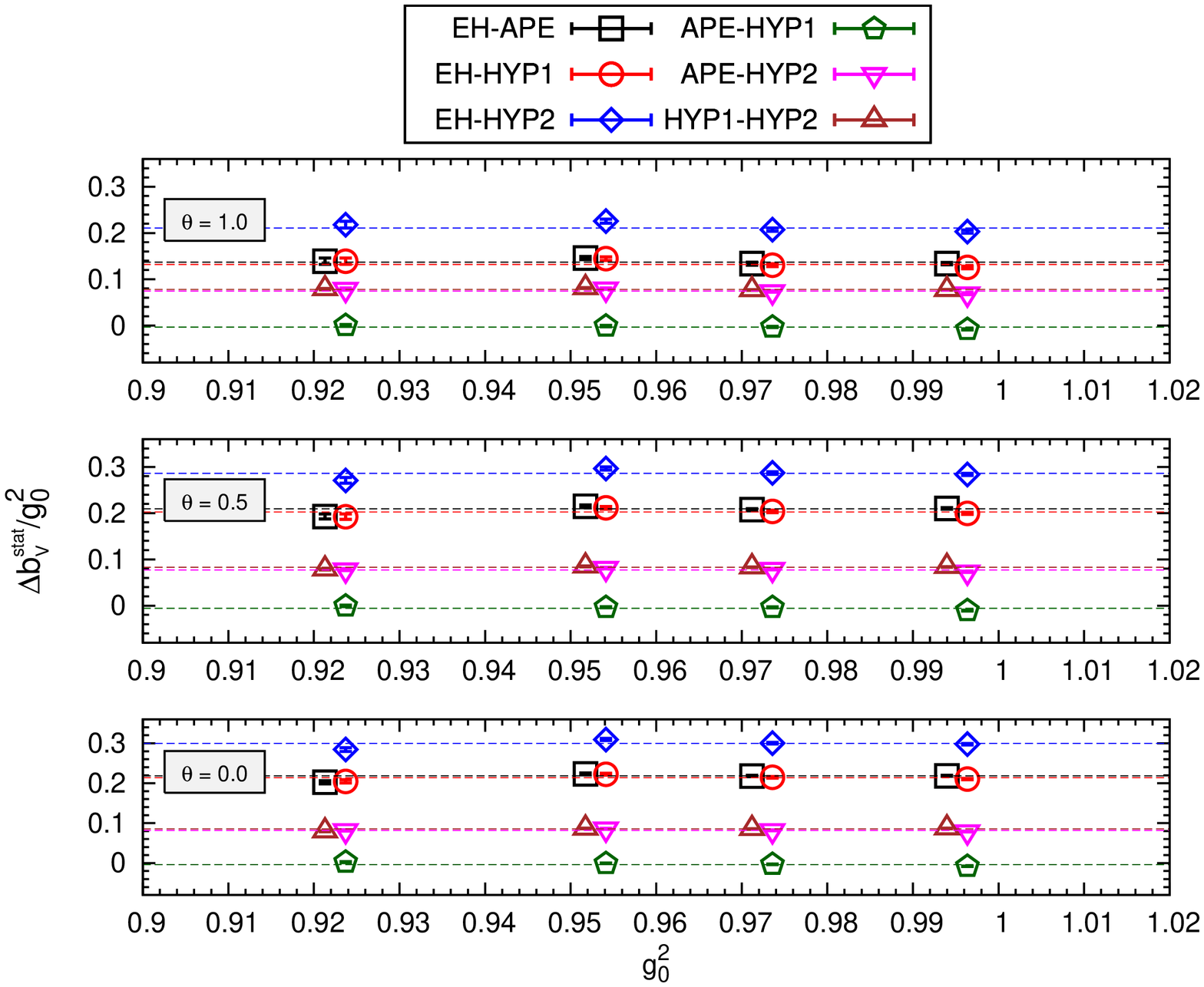, angle=-90, width=13.0 true cm}
\end{center}
\vskip -0.3cm \caption{$g_0^2$-dependence of $\Delta\bVstat/g_0^2$
corresponding to various combinations of the static actions and
for some choices of the $\theta$-angle. To improve the readability, 
points (squares and upper triangles) have been slightly shifted
along the horizontal axis. Dashed lines represent fits to a constant.
\label{fig:dbVstatnp}}
\end{figure}
\vskip -0.3cm
In order to get rid of the renormalization constants, we construct
the ratio of $M^\rI_{\rm\scriptscriptstyle R}$ at two different values
of the renormalized light-quark mass, i.e.
$Lm_{\rm\scriptscriptstyle R} = 0.24$ and $Lm_{\rm\scriptscriptstyle
R}=0$. This is not sufficient to isolate $\bVstat$, since the
improvement of the boundary light-quark source contains
 $b_\zeta$, which does not drop out in the ratio. Nevertheless,
$b_\zeta$ is independent of the static action. Therefore, it
cancels when we enforce the improvement condition that the ratio of
the three-point SF correlator be the same with two different static
actions ${\rm S}_1$ and ${\rm S}_2$ up to ${\rm O}(a^2)$-terms, i.e.
\begin{equation}
\label{eq:npimprconddb} \{1 + b^{\rm stat}_{\rm\scriptscriptstyle
V,S_1} a\mq\}\frac{M^\rI(T/2,m)}{M^\rI(T/2,m_{\rm
cr})}\biggl|_{\rm\scriptscriptstyle S_1} = \{1 + b^{\rm
stat}_{\rm\scriptscriptstyle V,S_2}
a\mq\}\frac{M^\rI(T/2,m)}{M^\rI(T/2,m_{\rm
cr})}\biggl|_{\rm\scriptscriptstyle S_2} \ + \ {\rm O}(a^2) \ .
\end{equation}
In the above equation, we decided to place the operator insertion in
the middle of the bulk and to choose $T/L=1$. This improvement
condition provides a non-perturbative  definition of $\Delta\bVstat
= b^{\rm stat}_{\rm\scriptscriptstyle V,S_1} - b^{\rm
stat}_{\rm\scriptscriptstyle V,S_2}$.
\parbreak
Simulations have been performed according to the parameters reported
in Table~\ref{tab:simpar}. In particular, the $\kappa$-values on the
right of the third column correspond to $Lm_{\rm\scriptscriptstyle
R}=0.24$, with $m_{\rm\scriptscriptstyle R}$ the quark mass 
renormalized in the SF scheme at scale $\mu= 1/(1.436r_0)$.  Numerical 
results of $\Delta\bVstat$ are reported in Table~\ref{tab:dbVstatnp} 
for the various independent combinations of the static actions and 
the usual $\theta$-angles. They are also represented in 
Fig.~\ref{fig:dbVstatnp}, where $\Delta\bVstat/g_0^2$ is plotted vs. 
$g_0^2$. A remarkable feature of the results is their flatness in 
$g_0^2$. Since
\begin{equation}
\bVstat = \frac{1}{2} + \bVstatone g_0^2 + O(g_0^4)\ ,
\end{equation}
this could be interpreted as a good signal of scaling and could
lead to the prompt conclusion that $\Delta\bVstat$ is
not dominated by ${\rm O}(g_0^4)$-terms. Nevertheless, we  
observe that $g_0^2$ varies from 0.924 to 0.996 in the range of the 
simulations, i.e. it changes by only 8\%. Such a small variation
could be well compatible with a slight change of the differences
$\Delta\bVstat$ even in a region not strictly close to the scaling one.
\parbreak
A second observation is that all the differences involving the EH
action have a significant dependence on $\theta$, with spreads
varying from 30\% to 60\% at the various bare gauge couplings. This
is a clear indication that large non-perturbative ${\rm O}(a)$
lattice artefacts affect our definition of $\bVstatEH$ 
based on the three-point SF correlator $M^\rI$. On the contrary, the 
remaining differences, involving exclusively the ALPHA actions, are 
much more universal in $\theta$: in these cases the spread among 
different definitions stays always below 0.01.
\parbreak
A fit of $\Delta\bVstat/g_0^2$ to a constant provides an effective
non-perturbative parametrization of the difference of the
improvement coefficients in the region of the Monte Carlo
simulations. Since we have no theoretical argument to privilege one
particular definition over the others, we decide to average the
results of the fits corresponding to the three $\theta$-values and
to assign the averages an absolute uncertainty as large as the maximal 
discrepancy between different $\theta$-determinations. In this way we obtain
\begin{alignat}{3}
& (\Delta\bVstat)_{\rm\scriptscriptstyle EH-APE}    & & \ = \ 0.19(8)g_0^2   \ , \label{eq:dbcheck} \\[2.0ex]
& (\Delta\bVstat)_{\rm\scriptscriptstyle EH-HYP1}   & & \ = \ 0.18(8)g_0^2   \ ,  \\[1.5ex]
& (\Delta\bVstat)_{\rm\scriptscriptstyle EH-HYP2}   & & \ = \ 0.27(9)g_0^2   \ ,  \\[1.5ex]
& (\Delta\bVstat)_{\rm\scriptscriptstyle APE-HYP1}  & & \ = \ -0.004(2)g_0^2 \ , \label{eq:dbapehypone} \\[1.5ex]
& (\Delta\bVstat)_{\rm\scriptscriptstyle APE-HYP2}  & & \ = \ 0.078(7)g_0^2  \ , \label{eq:dbapehyptwo} \\[1.5ex]
& (\Delta\bVstat)_{\rm\scriptscriptstyle HYP1-HYP2} & & \ = \ 0.082(7)g_0^2  \ .
\end{alignat}
\parbreak
In order to isolate the improvement coefficient $\bVstat$
corresponding to each static action, we perform an analytical
one-loop perturbative calculation of $\bVstatEH$ and $\bVstatAPE$.
To this aim, we expand the three-point SF correlator in powers of
$g_0^2$, i.e.
\begin{eqnarray}
M^\rI(x_0,m) & = & M^{(0)}(x_0,m^{(0)}) + \nonumber \\[1.5ex]
         & + & g_0^2\left[ M^{\rI(1)}\left(x_0,m^{(0)}\right) + m^{(1)}\partial_m M^{(0)}\left(x_0,m^{(0)}\right)
 + M^{\rI(1)}_{\rm\scriptscriptstyle b}\left(x_0,m^{(0)}\right) \right] + \nonumber \\[1.5ex]
         & + & {\rm O}(g_0^4) \ ,
\end{eqnarray}
where the perturbative coefficients of the bare quark mass $m^{(0)}$
and $m^{(1)}$ are chosen according to Eqs.~(3.29)-(3.30) of
\cite{Sint:1997jx}. Here $m_{\rm\scriptscriptstyle R}$ is defined as the renormalized quark mass
in the minimal subtraction scheme on the lattice at scale $\mu=1/L$. 
In this perturbative calculation we impose the improvement condition 
\begin{equation}
\label{eq:Mimprcond}
\frac{M^\rI_{\rm\scriptscriptstyle R}(T/2,0.24/L)}{M^\rI_{\rm\scriptscriptstyle R}(T/2,0)} = \ {\rm const.} \ + \ {\rm O}(a^2) \ ,
\end{equation}
\vskip 0.02cm
\begin{figure}[t]
\hskip 0.2cm
\epsfig{figure=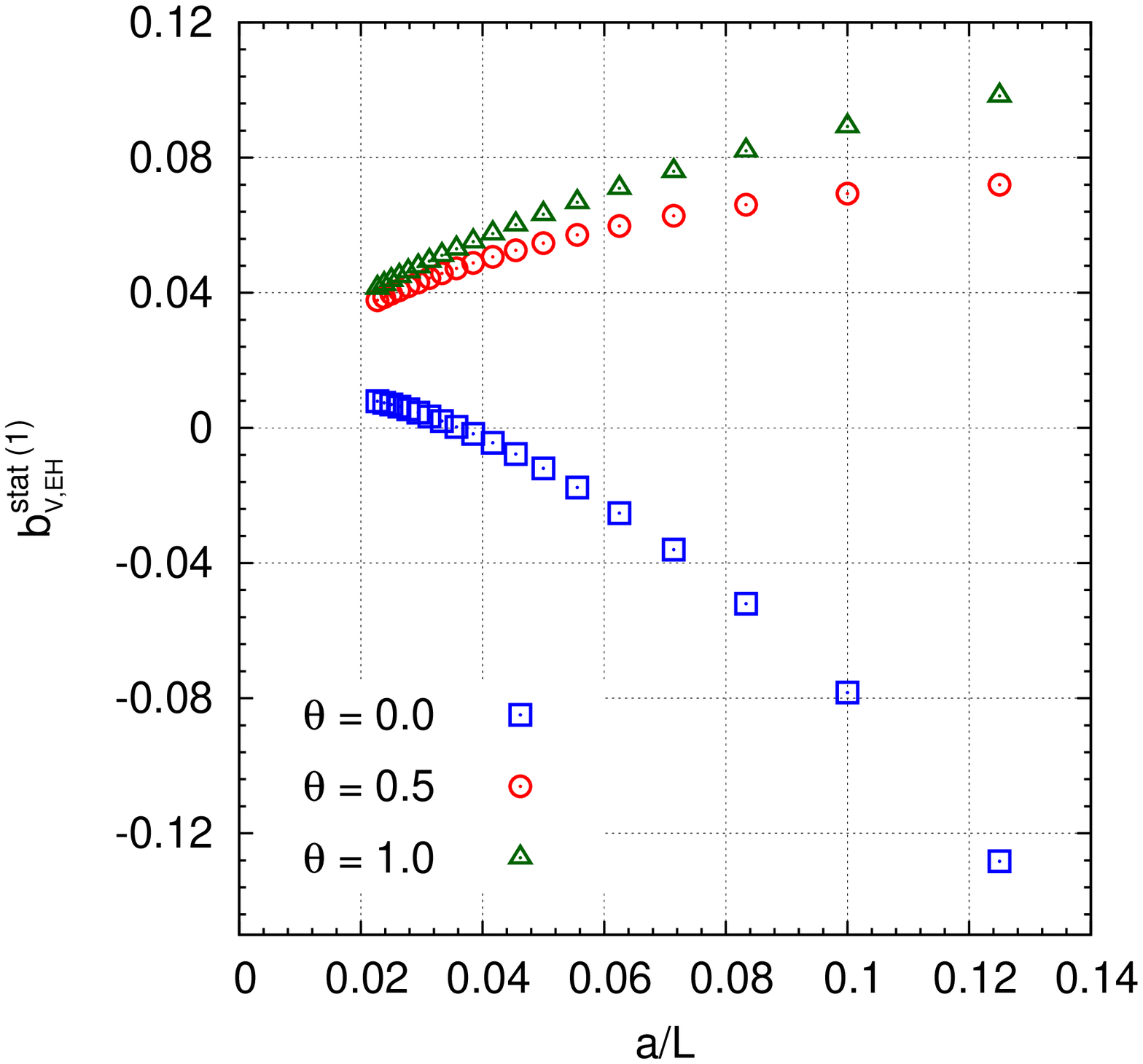, angle=-90, width=7 true cm}
\hskip 0.5cm
\epsfig{figure=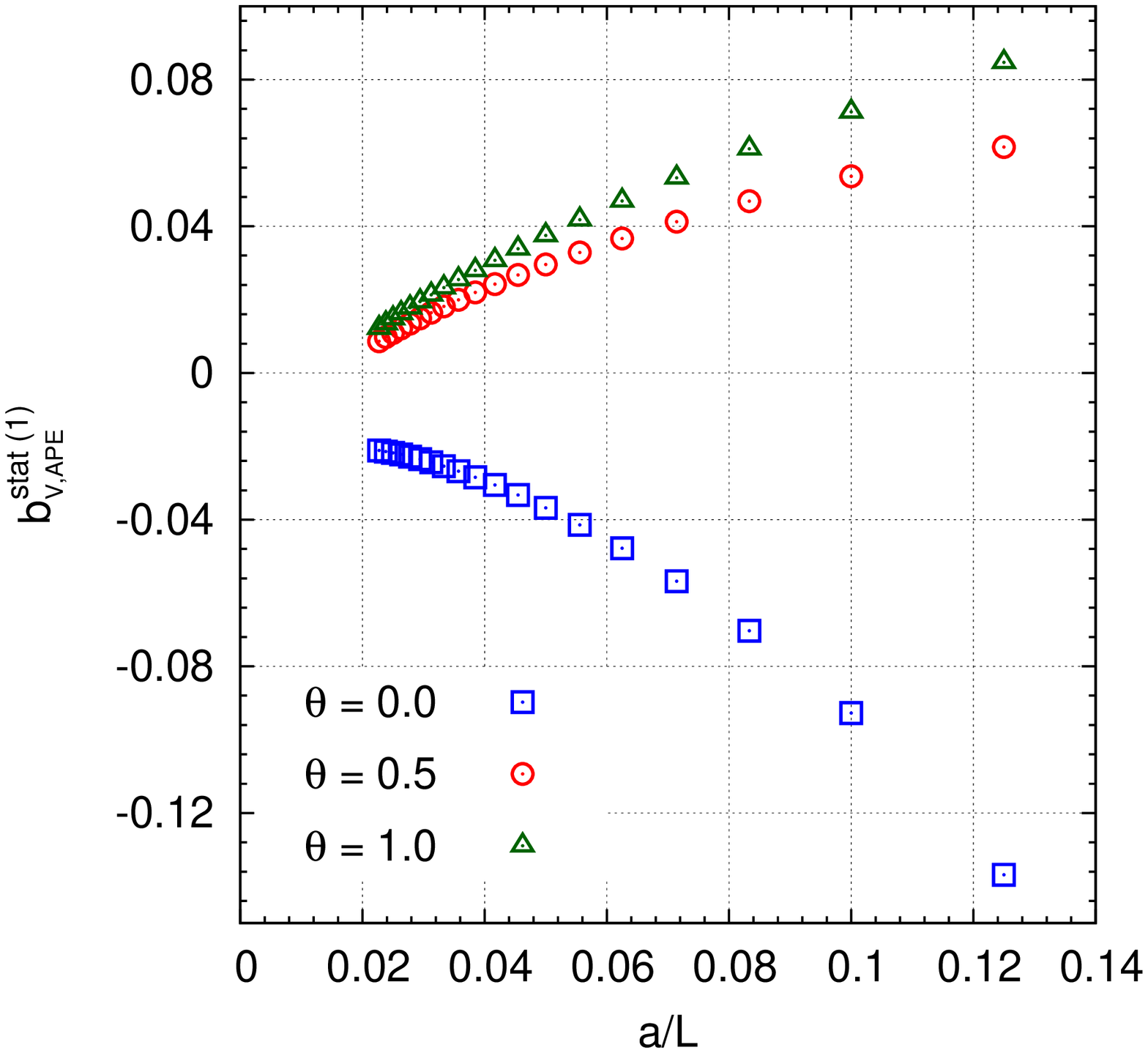, angle=-90, width=7 true cm}
\vskip -0.2cm
\caption{Continuum approach of $\bVstatone$ for the EH (left plot)
and APE (right plot) actions. Different choices of the
$\theta$-angle provide independent definitions of $\bVstatone$.
Plotted points correspond to $L/a=8,\dots,46$.\label{fig:bVstat1}}
\end{figure}
\noindent with aspect ratio $T/L=1$ and $\theta = 0.0,0.5,1.0$.
When expanded in perturbation theory, this equation 
provides for a definition of $\bVstatone + 3b_\zeta^{(1)}$ up to 
${\rm O}(a/L)$-terms. Since $b_\zeta^{(1)} = -0.06738(4)\times C_{\rm F}$ 
has been previously calculated in  \cite{Sint:1997jx}, this is 
sufficient to isolate
$\bVstatone$. 
\TABLE[r]{
  \small
  \centering
    \vbox{\vskip 0.8cm}
    \begin{tabular}{|c|c|c|}
      \hline
      & EH & APE \\
      \hline
      $\bVstatone$ & 0.013(1) & -0.018(1) \\
      \hline
    \end{tabular}
    \caption{$\bVstatone$ for the EH and APE actions.\label{tab:bvstatone}}
  }
\vskip 0.05cm
\noindent Lattice data are reported in
Tables~\ref{tab:bvstatEHone}-\ref{tab:bvstatAPEone} and plotted in
Fig.~\ref{fig:bVstat1}. Their continuum extrapolation leads to the
estimates quoted in Table~\ref{tab:bvstatone}. These correspond in
turn to an exact one-loop difference
\begin{equation}
\label{andiff}
(\Delta\bVstat)^{(1)}_{\rm\scriptscriptstyle EH-APE} = 0.0324(4)\ ,
\end{equation}
which is quite a bit off the central value of \Eq{eq:dbcheck}. 
Clearly, this difference may be attributed to the presence of 
non-negligible ${\rm O}(g_0^4)$-terms, which in principle could be there. 
However, the systematic uncertainty which characterizes the definition of 
the improvement coefficient with the EH action prevents us from making
a more precise statement. For this reason, we desist from quoting a final 
estimate of $\bVstatEH$. Instead, we use 
$b_{\rm\scriptscriptstyle V,APE}^{{\rm stat}(1)}$
to solve Eqs.~(\ref{eq:dbapehypone}-\ref{eq:dbapehyptwo}) and quote
\begin{alignat}{3}
& b_{\rm\scriptscriptstyle V,HYP1}^{ {\rm stat}} & \ \approx \ \frac{1}{2} -0.014(3)g_0^2 + {\rm O}(g_0^4) \ ,  \\[2.0ex]
& b_{\rm\scriptscriptstyle V,HYP2}^{ {\rm stat}} & \ \approx \ \frac{1}{2} -0.096(8)g_0^2 + {\rm O}(g_0^4) \ .
\end{alignat}
The reader should not be surprised to see that the difference 
$(\Delta\bVstat)^{(1)}_{\rm\scriptscriptstyle EH-APE}$ given in \Eq{andiff}
is more precise than the single values of $\bVstat$ reported in 
Table~\ref{tab:bvstatone}. Indeed, the continuum estimate of \Eq{andiff} 
has been obtained by extrapolating the difference of the lattice 
data reported in Tables~\ref{tab:bvstatEHone}-\ref{tab:bvstatAPEone}.
Part of the cutoff effects drops out in this difference, which makes the 
continuum extrapolation easier.

\subsection{A scaling test for $\cVstat$}

\begin{figure}[t]
\begin{center}
\hskip -1.0cm\epsfig{figure=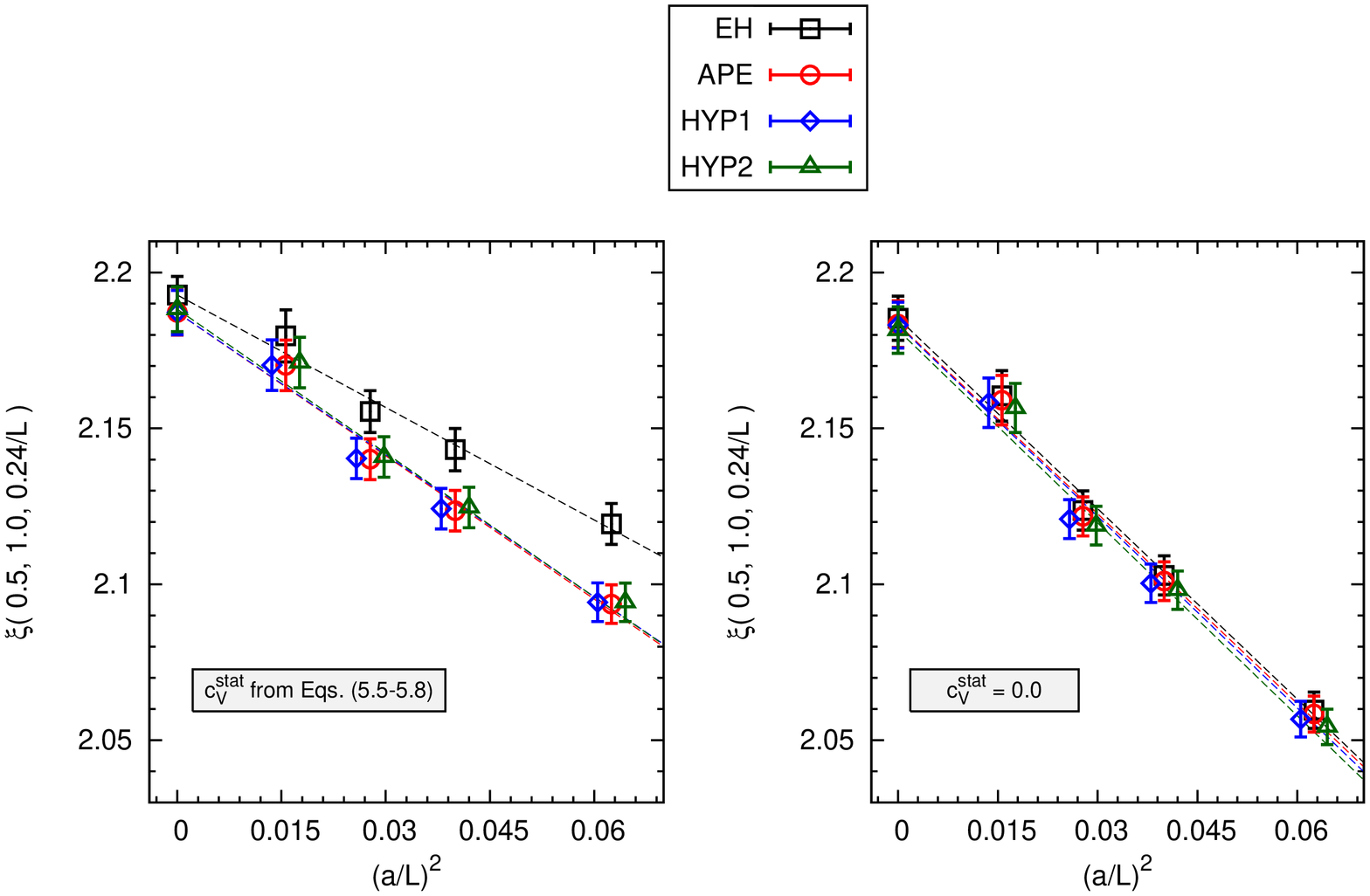, angle=-90, width=13.0 true cm}
\end{center}
\vskip -0.3cm \caption{Scaling plots for $\xi(0.5,1.0,0.24/L)$. To improve the readability, 
  some of the points (diamonds and triangles) have been slightly shifted along the 
  horizontal axis. Dashed curves represent independent linear fits in $(a/L)^2$. 
  Continuum extrapolated values are also shown.\label{fig:scaling}}
\end{figure}
Our non-perturbative data enable a scaling test of the three-point SF correlator 
$M^\rI$, useful to assess the effectiveness of our numerical 
determination of $\cVstat$. To this aim we introduce the ratio
\begin{equation}
\xi(\theta_1,\theta_2,m_{\rm\scriptscriptstyle R}) = \frac{M^\rI(T/2,m_{\rm\scriptscriptstyle R})|_{\theta_1}}
{M^\rI(T/2,m_{\rm\scriptscriptstyle R})|_{\theta_2}}\ , 
\end{equation}
which has a well defined continuum limit, 
with a theoretical rate of convergence proportional to 
${\rm O}(a^2)$ if the light-quark action is ${\rm O}(a)$-improved 
and $\cVstat$ is properly tuned. Fig.~\ref{fig:scaling} 
illustrates the approach of $\xi$ to the continuum, 
corresponding to the choice of parameters 
$\theta_1=0.5$, $\theta_2=1.0$ and $Lm_{\rm\scriptscriptstyle R}=0.24$. 
The left plot refers to the non-perturbative choice of 
$\cVstat$ provided by Eqs.~(\ref{eq:cvstatnp1}-\ref{eq:cvstatnp4});  
the right plot shows the unimproved case with $\cVstat=0$. Similar
plots are obtained with different $\theta$-angles and 
$m_{\rm\scriptscriptstyle R}$. 
\parbreak
We observe that all the static actions give comparable 
results and statistical uncertainties at finite lattice spacing, 
save for the EH one in the improved case. If we look at the right 
plot, we note that the total variation of $\xi$ in the simulation 
region is only about 5\%. This can be attributed to 
a significant cancellation of the ${\rm O}(a)$ lattice artefacts
between numerator and denominator, which on one hand gives
$\xi$ a good scaling behaviour also in absence of operator
improvement, but on the other makes it rather insensitive to
a change of $\cVstat$. Nevertheless, we find that once 
$\cVstat$ is switched on, the total variation of $\xi$ in the 
simulation region drops to 3\%, corresponding to a flatter 
approach to the continuum. As it might be expected, the strongest 
effect of $\cVstat$ is at $L/a=8$, where the central values of 
the lattice points are shifted by 1.5-2.9\%.


\section{Conclusions}

In this paper we have studied the renormalization of the static
vector current and its ${\rm O}(a)$-improvement in the quenched
approximation of lattice QCD. Quark degrees of freedom are described
by lattice Wilson-type fermions in the light sector and various
discretizations of the static fermions, including those introduced
some years ago by the ALPHA Collaboration (APE, HYP1, HYP2).
\parbreak
Owing to the chiral symmetry of the continuum theory, the RG running 
of the static vector and axial currents coincides. Since the latter 
has been extensively studied in the literature, a complete
description of the renormalization factor $\ZVstat$ is achieved
by simply fixing the ratio of the two renormalized  currents at a
given reference scale (in our study $\mu_{\rm ref}^{-1} = 2L_{\rm
max} = 1.436r_0$). To this aim we make use of an appropriate axial Ward
identity in the framework of the Schr\"odinger functional. The
enforcement of chiral symmetry up to ${\rm O}(a^2)$-terms provides
us with a lever to tune the improvement coefficient $\cVstat$.
Unfortunately, the resulting determination is not fully
non-perturbative, since it relies upon a previous computation of
$\cAstat$ which is only effective, i.e. correct up to ${\rm
O}(g_0^4)$-terms. With regard to the numerical results, a comparison
of the Monte Carlo simulations with a one-loop perturbative
calculation shows that large higher-order contributions affect
$\cVstat$ within the explored region of the gauge coupling $(6.0\le
\beta \le 6.5)$. On the other hand, we observe a good agreement
between the non-perturbative determination of the ${\rm
O}(a)$-improved ratio $\ZAstat/\ZVstat$ and its one-loop approximation.
\parbreak
The ${\cal O}(a)$-improvement programme is carried out at non-zero
light-quark mass via the introduction of a second improvement
coefficient $\bVstat$. This is tuned on the basis of an independent
condition involving a boundary-to-boundary three-point correlator of
the static vector current, out of the chiral limit. The coefficient 
$\bVstat$ is  studied at one-loop order in perturbation theory for the 
EH and APE actions. To extend our study to the HYP actions, where
perturbation theory is not easily handled, we adopt a mixed
strategy: the difference $\Delta\bVstat$ of the improvement
coefficients between two different static discretizations is computed
non-perturbatively and the one-loop estimate with the APE action is
used to isolate $\bVstat$ in the HYP1 and HYP2 cases up to ${\rm
O}(g_0^4)$-terms.
It has to be said that a direct comparison of the non-perturbative 
estimate of $(\Delta\bVstat)_{\scriptscriptstyle\rm EH-APE}$ 
with its one-loop value shows that the amount of such 
${\rm O}(g_0^4)$-terms could be non-negligible and hard to control.
Nevertheless, this problem seems to characterize the EH fermions
more than their statistically improved versions, for which a better
agreement with perturbation theory is expected on the basis of the
experience gathered by the ALPHA Collaboration in previous studies
of the static axial current. 

Anyway, one should always keep in mind that $\bVstat$ enters the 
improved static vector current accompanied by a factor of $am_{\rm q}$, 
which is rather small at light-quark masses up to the strange one and 
the commonly affordable lattice spacings. In this sense, it 
constitutes a subdominant contribution, which is not expected to 
have a crucial effect on the scaling behaviour of phenomenological 
matrix elements of the static vector current with external 
$B_d$- or $B_s$-meson states.

\section*{Acknowledgments}
A special thanks goes to R.~Sommer for invaluable support during all the stages
of this work. We also thank D.~Guazzini, M.~Della Morte, M.~Papinutto, C.~Pena and 
H.~Wittig for useful discussions. Partial financial support from the 
Alexander-von-Humboldt Stiftung is acknowledged. We acknowledge DESY Hamburg and 
the Institut f\"ur Kernphysik - Universit\"at Mainz for providing hospitality 
during the intermediate stage of the project, as well as the computing centre of 
DESY Zeuthen for its technical support. This work was supported in part by the EU 
Contract No. MRTN-CT-2006-035482, ``FLAVIAnet''.

\appendix

\newpage

\section{Additional tables}

\begin{table}[!ht]
  \small
    \begin{center}
      \vbox{\vskip 0.2cm}
      \begin{tabular}{cccc}
        \hline\hline\\[-2.0ex]
        $L/a$ & $\cVstatoneEH|_{(\theta_1,\theta_2)=(0.0,0.5)}$ & $\cVstatoneEH|_{(\theta_1,\theta_2)=(0.0,1.0)}$ & $\cVstatoneEH|_{(\theta_1,\theta_2)=(0.5,1.0)}$  \\ \\[-2.0ex]
        \hline\\[-2.0ex]
        4  & $-1.21318905506\times 10^{-1}$ &  $-5.56860354782\times 10^{-2}$ &  $-1.14115660920\times 10^{-2}$ \\
        6  & $-8.05799533807\times 10^{-2}$ &  $-3.56181870915\times 10^{-2}$ &  $-7.37458670533\times 10^{-3}$ \\
        8  & $-5.45943895816\times 10^{-2}$ &  $-2.14296960443\times 10^{-2}$ &  $-1.11423236142\times 10^{-3}$ \\
        10 & $-4.17792968218\times 10^{-2}$ &  $-1.58212044566\times 10^{-2}$ &  $-1.04666543337\times 10^{-4}$ \\
        12 & $-3.39132673910\times 10^{-2}$ &  $-1.27133821723\times 10^{-2}$ &  $ 4.11644560339\times 10^{-5}$ \\
        14 & $-2.84872496446\times 10^{-2}$ &  $-1.06310641447\times 10^{-2}$ &  $ 7.08421808901\times 10^{-5}$ \\
        16 & $-2.44775851326\times 10^{-2}$ &  $-9.08817318086\times 10^{-3}$ &  $ 1.12471960839\times 10^{-4}$ \\
        18 & $-2.13776226148\times 10^{-2}$ &  $-7.87813705828\times 10^{-3}$ &  $ 1.78912840345\times 10^{-4}$ \\
        20 & $-1.89020434292\times 10^{-2}$ &  $-6.89431955817\times 10^{-3}$ &  $ 2.63691059080\times 10^{-4}$ \\
        22 & $-1.68756672083\times 10^{-2}$ &  $-6.07394710919\times 10^{-3}$ &  $ 3.59366131023\times 10^{-4}$ \\
        24 & $-1.51841738803\times 10^{-2}$ &  $-5.37672581715\times 10^{-3}$ &  $ 4.60431552309\times 10^{-4}$ \\
        26 & $-1.37494665187\times 10^{-2}$ &  $-4.77520683457\times 10^{-3}$ &  $ 5.63221867541\times 10^{-4}$ \\
        28 & $-1.25162477343\times 10^{-2}$ &  $-4.24986678175\times 10^{-3}$ &  $ 6.65402844941\times 10^{-4}$ \\
        30 & $-1.14441713189\times 10^{-2}$ &  $-3.78634326450\times 10^{-3}$ &  $ 7.65528806751\times 10^{-4}$ \\
        32 & $-1.05031230408\times 10^{-2}$ &  $-3.37380171813\times 10^{-3}$ &  $ 8.62744511694\times 10^{-4}$ \\ \\[-2.0ex]
        \hline\hline
      \end{tabular}
      \caption{Three different determinations of the one-loop contribution to $\cVstat$ with EH static fermions according to the improvement condition \protect\Eq{Rimprcond}. Numbers refer to the topology $\cT=2$.\label{tab:cvstatEHone}}
    \end{center}
    \vskip -2.0cm
  \small
    \begin{center}
      \vbox{\vskip 0.2cm}
      \begin{tabular}{cccc}
        \hline\hline\\[-2.0ex]
        $L/a$ & $\cVstatoneAPE|_{(\theta_1,\theta_2)=(0.0,0.5)}$ & $\cVstatoneAPE|_{(\theta_1,\theta_2)=(0.0,1.0)}$ & $\cVstatoneAPE|_{(\theta_1,\theta_2)=(0.5,1.0)}$  \\ \\[-2.0ex]
        \hline\\[-2.0ex]
        4  & $-1.22686673290\times 10^{-1}$  & $-6.58409712344\times 10^{-2}$  & $-2.74941292579\times 10^{-2}$  \\
        6  & $-7.75983902179\times 10^{-2}$  & $-3.92522157408\times 10^{-2}$  & $-1.51643269042\times 10^{-2}$  \\
        8  & $-4.86871345865\times 10^{-2}$  & $-2.02408469320\times 10^{-2}$  & $-2.81570404300\times 10^{-3}$  \\
        10 & $-3.41205392729\times 10^{-2}$  & $-1.17290094210\times 10^{-2}$  & $ 1.82812424278\times 10^{-3}$  \\
        12 & $-2.51362502419\times 10^{-2}$  & $-6.77983138946\times 10^{-3}$  & $ 4.26399256380\times 10^{-3}$  \\
        14 & $-1.89433155853\times 10^{-2}$  & $-3.44517803597\times 10^{-3}$  & $ 5.84345845792\times 10^{-3}$  \\
        16 & $-1.43758718224\times 10^{-2}$  & $-9.98819307355\times 10^{-4}$  & $ 6.99872560335\times 10^{-3}$  \\
        18 & $-1.08514987939\times 10^{-2}$  & $ 8.93569798294\times 10^{-4}$  & $ 7.90351127450\times 10^{-3}$  \\
        20 & $-8.04166342915\times 10^{-3}$  & $ 2.41124455308\times 10^{-3}$  & $ 8.64240269108\times 10^{-3}$  \\
        22 & $-5.74490428783\times 10^{-3}$  & $ 3.66095340524\times 10^{-3}$  & $ 9.26291571099\times 10^{-3}$  \\
        24 & $-3.83000893356\times 10^{-3}$  & $ 4.71111391836\times 10^{-3}$  & $ 9.79458495765\times 10^{-3}$  \\
        26 & $-2.20751253184\times 10^{-3}$  & $ 5.60799879909\times 10^{-3}$  & $ 1.02571343800\times 10^{-2}$  \\
        28 & $-8.14172229101\times 10^{-4}$  & $ 6.38421491915\times 10^{-3}$  & $ 1.06644451570\times 10^{-2}$  \\
        30 & $ 3.96087085934\times 10^{-4}$  & $ 7.06350798921\times 10^{-3}$  & $ 1.10266744094\times 10^{-2}$  \\
        32 & $ 1.45762004875\times 10^{-3}$  & $ 7.66362155507\times 10^{-3}$  & $ 1.13514918436\times 10^{-2}$  \\ \\[-2.0ex]
        \hline\hline
      \end{tabular}
      \caption{Three different determinations of the one-loop contribution to $\cVstat$ with APE static fermions according to the improvement condition \protect\Eq{Rimprcond}. Numbers refer to the topology $\cT=2$.\label{tab:cvstatAPEone}}
    \end{center}
\end{table}

\newpage

\begin{table}[!ht]
  \small
    \begin{center}
      \vbox{\vskip 0.2cm}
      \begin{tabular}{cccc}
        \hline\hline\\[-2.0ex]
        $L/a$ & $\cR^{(1)}_{\rm\scriptscriptstyle EH}(\theta=0.0)$ & $\cR^{(1)}_{\rm\scriptscriptstyle EH}(\theta=0.5)$ & $\cR^{(1)}_{\rm\scriptscriptstyle EH}(\theta=1.0)$  \\ \\[-2.0ex]
        \hline\\[-2.0ex]
        4  &  $3.30112791641\times 10^{-2}$  &  $5.07890218387\times 10^{-2}$  &  $5.32679355672\times 10^{-2}$ \\
        6  &  $3.95982301495\times 10^{-2}$  &  $4.76144903008\times 10^{-2}$  &  $4.87823906356\times 10^{-2}$ \\
        8  &  $5.12452416492\times 10^{-2}$  &  $5.53446719297\times 10^{-2}$  &  $5.54812561435\times 10^{-2}$ \\
        10 &  $5.62747608283\times 10^{-2}$  &  $5.87919096424\times 10^{-2}$  &  $5.88023249365\times 10^{-2}$ \\
        12 &  $5.88913680345\times 10^{-2}$  &  $6.05967926872\times 10^{-2}$  &  $6.05933519321\times 10^{-2}$ \\
        14 &  $6.04217811204\times 10^{-2}$  &  $6.16508787417\times 10^{-2}$  &  $6.16457789162\times 10^{-2}$ \\
        16 &  $6.13931315301\times 10^{-2}$  &  $6.23177973609\times 10^{-2}$  &  $6.23106907219\times 10^{-2}$ \\
        18 &  $6.20478428063\times 10^{-2}$  &  $6.27659838840\times 10^{-2}$  &  $6.27559138013\times 10^{-2}$ \\
        20 &  $6.25099352992\times 10^{-2}$  &  $6.30815917333\times 10^{-2}$  &  $6.30682137670\times 10^{-2}$ \\
        22 &  $6.28481782065\times 10^{-2}$  &  $6.33122587177\times 10^{-2}$  &  $6.32956656095\times 10^{-2}$ \\
        24 &  $6.31031850685\times 10^{-2}$  &  $6.34860186764\times 10^{-2}$  &  $6.34665140076\times 10^{-2}$ \\
        26 &  $6.33001903502\times 10^{-2}$  &  $6.36202279760\times 10^{-2}$  &  $6.35981895849\times 10^{-2}$ \\
        28 &  $6.34555419363\times 10^{-2}$  &  $6.37260940915\times 10^{-2}$  &  $6.37019044423\times 10^{-2}$ \\
        30 &  $6.35802087156\times 10^{-2}$  &  $6.38111147867\times 10^{-2}$  &  $6.37851294648\times 10^{-2}$ \\
        32 &  $6.36817718015\times 10^{-2}$  &  $6.38804596546\times 10^{-2}$  &  $6.38529951792\times 10^{-2}$ \\ \\[-2.0ex]
        \hline\hline
      \end{tabular}
      \caption{Three different determinations of the one-loop contribution to the ${\rm O}(a)$-improved WI with EH static fermions. Numbers refer to the topology $\cT=2$.\label{tab:REHone}}
    \end{center}
    \vskip -2.0cm
  \small
    \begin{center}
      \vbox{\vskip 0.2cm}
      \begin{tabular}{cccc}
        \hline\hline\\[-2.0ex]
        $L/a$ & $\cR^{(1)}_{\rm\scriptscriptstyle APE}(\theta=0.0)$ & $\cR^{(1)}_{\rm\scriptscriptstyle APE}(\theta=0.5)$ & $\cR^{(1)}_{\rm\scriptscriptstyle APE}(\theta=1.0)$  \\ \\[-2.0ex]
        \hline\\[-2.0ex]
        4  & $7.83613664415\times 10^{-2}$  & $9.63395380897\times 10^{-2}$  & $1.02312037738\times 10^{-1}$  \\
        6  & $8.29999758507\times 10^{-2}$  & $9.07196239420\times 10^{-2}$  & $9.31211717407\times 10^{-2}$  \\
        8  & $9.43090648772\times 10^{-2}$  & $9.79649261355\times 10^{-2}$  & $9.83100792082\times 10^{-1}$  \\
        10 & $9.92568717719\times 10^{-2}$  & $1.01312590309\times 10^{-1}$  & $1.01130674952\times 10^{-1}$  \\
        12 & $1.01847718258\times 10^{-1}$  & $1.03111765757\times 10^{-1}$  & $1.02755357444\times 10^{-1}$  \\
        14 & $1.03368358402\times 10^{-1}$  & $1.04185677884\times 10^{-1}$  & $1.03765015807\times 10^{-1}$  \\
        16 & $1.04335470132\times 10^{-1}$  & $1.04878533404\times 10^{-1}$  & $1.04436312823\times 10^{-1}$  \\
        18 & $1.04988139034\times 10^{-1}$  & $1.05352674754\times 10^{-1}$  & $1.04907826792\times 10^{-1}$  \\
        20 & $1.05449158322\times 10^{-1}$  & $1.05692363172\times 10^{-1}$  & $1.05253903968\times 10^{-1}$  \\
        22 & $1.05786795092\times 10^{-1}$  & $1.05944779845\times 10^{-1}$  & $1.05517080769\times 10^{-1}$  \\
        24 & $1.06041438462\times 10^{-1}$  & $1.06138003224\times 10^{-1}$  & $1.05723087837\times 10^{-1}$  \\
        26 & $1.06238214652\times 10^{-1}$  & $1.06289597523\times 10^{-1}$  & $1.05888244607\times 10^{-1}$  \\
        28 & $1.06393415756\times 10^{-1}$  & $1.06411014965\times 10^{-1}$  & $1.06023326229\times 10^{-1}$  \\
        30 & $1.06517980204\times 10^{-1}$  & $1.06509988458\times 10^{-1}$  & $1.06135695966\times 10^{-1}$  \\
        32 & $1.06619471630\times 10^{-1}$  & $1.06591897793\times 10^{-1}$  & $1.06230536156\times 10^{-1}$  \\ \\[-2.0ex]
        \hline\hline
      \end{tabular}
      \caption{Three different determinations of the one-loop contribution to the ${\rm O}(a)$-improved WI with APE static fermions. Numbers refer to the topology $\cT=2$.\label{tab:RAPEone}}
    \end{center}
\end{table}

\newpage

\begin{table}[!ht]
  \small
    \begin{center}
      \vbox{\vskip 0.2cm}
      \begin{tabular}{cccc}
        \hline\hline\\[-2.0ex]
        $L/a$ & $\bVstatoneEH(\theta=0.0)$ & $\bVstatoneEH(\theta=0.5)$ & $\bVstatoneEH(\theta=1.0)$  \\ \\[-2.0ex]
        \hline\\[-2.0ex]
        6  & $-1.92494990552\times 10^{-1}$  &  $7.59451390152\times 10^{-2}$ &  $7.94433253811\times 10^{-2}$ \\
        8  & $-1.28284962759\times 10^{-1}$  &  $7.19668221779\times 10^{-2}$ &  $9.82580984223\times 10^{-2}$ \\
        10 & $-7.83097066265\times 10^{-2}$  &  $6.93670764950\times 10^{-2}$ &  $8.92282752595\times 10^{-2}$ \\
        12 & $-5.20449399705\times 10^{-2}$  &  $6.60882882658\times 10^{-2}$ &  $8.20217785512\times 10^{-2}$ \\
        14 & $-3.59939921230\times 10^{-2}$  &  $6.27858498787\times 10^{-2}$ &  $7.59502707254\times 10^{-2}$ \\
        16 & $-2.52419762755\times 10^{-2}$  &  $5.97829596381\times 10^{-2}$ &  $7.09345102910\times 10^{-2}$ \\
        18 & $-1.76143672046\times 10^{-2}$  &  $5.71059870738\times 10^{-2}$ &  $6.67548992369\times 10^{-2}$ \\
        20 & $-1.19885067448\times 10^{-2}$  &  $5.47150817151\times 10^{-2}$ &  $6.32117968093\times 10^{-2}$ \\
        22 & $-7.71840193796\times 10^{-3}$  &  $5.25663880980\times 10^{-2}$ &  $6.01574097982\times 10^{-2}$ \\
        24 & $-4.40467430491\times 10^{-3}$  &  $5.06225291197\times 10^{-2}$ &  $5.74858429769\times 10^{-2}$ \\
        26 & $-1.78765598057\times 10^{-3}$  &  $4.88528309736\times 10^{-2}$ &  $5.51202560392\times 10^{-2}$ \\
        28 & $ 3.08255934246\times 10^{-4}$  &  $4.72321849465\times 10^{-2}$ &  $5.30036608571\times 10^{-2}$ \\
        30 & $ 2.00560511473\times 10^{-3}$  &  $4.57399574759\times 10^{-2}$ &  $5.10928394946\times 10^{-2}$ \\
        32 & $ 3.39251992852\times 10^{-3}$  &  $4.43591299385\times 10^{-2}$ &  $4.93543727896\times 10^{-2}$ \\
        34 & $ 4.53369168047\times 10^{-3}$  &  $4.30755529827\times 10^{-2}$ &  $4.77620635919\times 10^{-2}$ \\ \\[-2.0ex]
        \hline\hline
      \end{tabular}
      \caption{Three different determinations of the one-loop contribution to $\bVstat$ with EH static fermions according to the improvement condition \protect\Eq{eq:Mimprcond}.\label{tab:bvstatEHone}}
    \end{center}
    \vskip -2.0cm
  \small
    \begin{center}
      \vbox{\vskip 0.2cm}
      \begin{tabular}{cccc}
        \hline\hline\\[-2.0ex]
        $L/a$ & $\bVstatoneAPE(\theta=0.0)$ & $\bVstatoneAPE(\theta=0.5)$ & $\bVstatoneAPE(\theta=1.0)$  \\ \\[-2.0ex]
        \hline\\[-2.0ex]
        6  &  $-1.92340408686\times 10^{-1}$  & $7.38125998350\times 10^{-2}$  &  $7.32869781736\times 10^{-2}$ \\
        8  &  $-1.36845655003\times 10^{-1}$  & $6.16212128505\times 10^{-2}$  &  $8.47024088104\times 10^{-2}$ \\
        10 &  $-9.27641005323\times 10^{-2}$  & $5.36113971906\times 10^{-2}$  &  $7.11683168091\times 10^{-2}$ \\
        12 &  $-7.03192609445\times 10^{-2}$  & $4.68591075301\times 10^{-2}$  &  $6.11505675084\times 10^{-2}$ \\
        14 &  $-5.67994367873\times 10^{-2}$  & $4.12586767144\times 10^{-2}$  &  $5.32267457567\times 10^{-2}$ \\
        16 &  $-4.77974262206\times 10^{-2}$  & $3.66636780399\times 10^{-2}$  &  $4.69184320708\times 10^{-2}$ \\
        18 &  $-4.14377364146\times 10^{-2}$  & $3.28289740025\times 10^{-2}$  &  $4.17877409334\times 10^{-2}$ \\
        20 &  $-3.67710201232\times 10^{-2}$  & $2.95586046439\times 10^{-2}$  &  $3.75125941270\times 10^{-2}$ \\
        22 &  $-3.32534218174\times 10^{-2}$  & $2.67170066138\times 10^{-2}$  &  $3.38739675354\times 10^{-2}$ \\
        24 &  $-3.05477831878\times 10^{-2}$  & $2.42109755463\times 10^{-2}$  &  $3.07226742530\times 10^{-2}$ \\
        26 &  $-2.84338924925\times 10^{-2}$  & $2.19744102751\times 10^{-2}$  &  $2.79543183872\times 10^{-2}$ \\
        28 &  $-2.67622184958\times 10^{-2}$  & $1.99587562440\times 10^{-2}$  &  $2.54935704579\times 10^{-2}$ \\
        30 &  $-2.54281229985\times 10^{-2}$  & $1.81272438403\times 10^{-2}$  &  $2.32844980335\times 10^{-2}$ \\
        32 &  $-2.43562070602\times 10^{-2}$  & $1.64513560157\times 10^{-2}$  &  $2.12845506743\times 10^{-2}$ \\
        34 &  $-2.34911031398\times 10^{-2}$  & $1.49085018459\times 10^{-2}$  &  $1.94607725049\times 10^{-2}$ \\ \\[-2.0ex]
        \hline\hline
      \end{tabular}
      \caption{Three different determinations of the one-loop contribution to $\bVstat$ with APE static fermions according to the improvement condition \protect\Eq{eq:Mimprcond}.\label{tab:bvstatAPEone}}
    \end{center}
\end{table}

\newpage

\begin{table}[!ht]
  \small
    \begin{center}
      \vbox{\vskip 1.2cm}
      \begin{tabular}{cccccc}
        \hline\hline\\[-2.0ex]
        $(\theta_1,\theta_2)$ & $\beta$ & $\cVstatEH$ & $\cVstatAPE$ & $\cVstatHYPone$ & $\cVstatHYPtwo$ \\ \\[-2.0ex]
        \hline\\[-2.0ex]
        (0.0,0.5) & 6.0219 & 0.756(22) & 0.478(18) & 0.513(18) & 0.553(18) \\
                  & 6.1628 & 0.577(18) & 0.337(14) & 0.360(14) & 0.409(14) \\
                  & 6.2885 & 0.548(17) & 0.334(14) & 0.359(14) & 0.416(14) \\
                  & 6.4956 & 0.399(18) & 0.240(14) & 0.261(14) & 0.324(14) \\ \\[-2.0ex]
        \hline\\[-2.0ex]
        (0.0,1.0) & 6.0219 & 0.707(10) & 0.433(8)  & 0.467(8)  & 0.508(8)  \\
                  & 6.1628 & 0.566(8)  & 0.328(6)  & 0.352(7)  & 0.402(7)  \\
                  & 6.2885 & 0.532(8)  & 0.318(6)  & 0.342(6)  & 0.398(7)  \\
                  & 6.4956 & 0.406(8)  & 0.242(7)  & 0.263(7)  & 0.327(7)  \\ \\[-2.0ex]
        \hline\\[-2.0ex]
        (0.5,1.0) & 6.0219 & 0.690(8)  & 0.419(6)  & 0.452(7)  & 0.493(6)  \\
                  & 6.1628 & 0.562(6)  & 0.325(5)  & 0.349(5)  & 0.399(5)  \\
                  & 6.2885 & 0.527(6)  & 0.312(5)  & 0.336(5)  & 0.392(5)  \\
                  & 6.4956 & 0.409(6)  & 0.242(5)  & 0.264(5)  & 0.328(5)  \\ \\[-2.0ex]
        \hline\hline
      \end{tabular}
      \caption{Non-perturbative determinations of $\cVstat$ for various gauge couplings and static actions. Different choices of $(\theta_1,\theta_2)$
        correspond to independent definitions of the improvement coefficient.\label{tab:cvstatnp}}
    \end{center}
  \small
    \begin{center}
      \vbox{\vskip 1.5cm}
      \begin{tabular}{cccccc}
        \hline\hline\\[-2.0ex]
        $\theta$ & $\beta$ & $[\ZAstat/\ZVstat]_{\rm\scriptscriptstyle EH}$ & $[\ZAstat/\ZVstat]_{\rm\scriptscriptstyle APE}$ & $[\ZAstat/\ZVstat]_{\rm\scriptscriptstyle HYP1}$ & $[\ZAstat/\ZVstat]_{\rm\scriptscriptstyle HYP2}$ \\ \\[-2.0ex]
        \hline\\[-2.0ex]
        0.0 & 6.0219 &  0.9549(13) & 0.9611(13) & 0.9662(13) & 0.9643(13) \\
            & 6.1628 &  0.9564(10) & 0.9725(10) & 0.9771(9)  & 0.9785(10) \\
            & 6.2885 &  0.9585(8)  & 0.9799(9)  & 0.9837(9)  & 0.9859(10) \\
            & 6.4956 &  0.9527(7)  & 0.9823(7)  & 0.9847(7)  & 0.9890(6)  \\ \\[-2.0ex]
        \hline\\[-2.0ex]
        0.5 & 6.0219 &  0.9549(13) & 0.9601(12) & 0.9651(12) & 0.9633(12) \\
            & 6.1628 &  0.9562(10) & 0.9723(10) & 0.9769(10) & 0.9784(9)  \\
            & 6.2885 &  0.9582(8)  & 0.9796(8)  & 0.9834(7)  & 0.9856(9)  \\
            & 6.4956 &  0.9528(6)  & 0.9823(7)  & 0.9847(6)  & 0.9890(7)  \\ \\[-2.0ex]
        \hline\\[-2.0ex]
        1.0 & 6.0219 &  0.9540(11) & 0.9601(11) & 0.9651(11) & 0.9633(11) \\
            & 6.1628 &  0.9561(9)  & 0.9723(10) & 0.9769(10) & 0.9784(8)  \\
            & 6.2885 &  0.9583(8)  & 0.9796(7)  & 0.9834(7)  & 0.9856(9)  \\
            & 6.4956 &  0.9528(6)  & 0.9823(6)  & 0.9847(5)  & 0.9890(6)  \\ \\[-2.0ex]
        \hline\hline
      \end{tabular}
      \caption{Non-perturbative determinations of the ${\rm O}(a)$-improved ratio $\ZAstat/\ZVstat$ for various gauge couplings and
        static actions. Different choices of $\theta$ correspond to independent definitions of the WI.\label{tab:zazvnp}}
    \end{center}
\end{table}

\begin{table}[!ht]
  \small
    \begin{center}
      \vbox{\vskip 0.0cm}
      \begin{tabular}{ccccc}
        \hline\hline
        $\theta$ & $\beta$ & $(\Delta\bVstat)_{\rm\scriptscriptstyle\ EH-APE\ }$  & $(\Delta\bVstat)_{\rm\scriptscriptstyle\ EH-HYP1\ }$ & $(\Delta\bVstat)_{\rm\scriptscriptstyle\ EH-HYP2\ }$ \\
        \hline\\[-2.0ex]
        0.0 & 6.0219 &  0.2177(14) &  0.2098(15) &  0.2967(18) \\
            & 6.1628 &  0.2122(17) &  0.2091(18) &  0.2925(21) \\
            & 6.2885 &  0.2130(22) &  0.2123(22) &  0.2953(26) \\
            & 6.4956 &  0.1869(37) &  0.1890(39) &  0.2626(11) \\ \\[-2.0ex]
        \hline\hline
      \end{tabular}
      \vskip 0.3cm
      \begin{tabular}{ccccc}
        \hline\hline
        $\theta$ & $\beta$ & $(\Delta\bVstat)_{\rm\scriptscriptstyle APE-HYP1}$ & $(\Delta\bVstat)_{\rm\scriptscriptstyle APE-HYP2}$ & $(\Delta\bVstat)_{\rm\scriptscriptstyle HYP1-HYP2}$ \\
        \hline\\[-2.0ex]
        0.0 & 6.0219 & -0.0079(7)  &  0.0787(10) &  0.0865(6)   \\
            & 6.1628 & -0.0031(7)  &  0.0800(10) &  0.0831(6)   \\
            & 6.2885 & -0.0008(8)  &  0.0820(12) &  0.0828(7)   \\
            & 6.4956 & 0.0021(11)  &  0.0756(17) &  0.0735(10)  \\ \\[-2.0ex]
        \hline\hline
      \end{tabular}
    \end{center}
  \small
    \begin{center}
      \vbox{\vskip 0.6cm}
      \begin{tabular}{ccccc}
        \hline\hline
        $\theta$ & $\beta$ & $(\Delta\bVstat)_{\rm\scriptscriptstyle\ EH-APE\ }$  & $(\Delta\bVstat)_{\rm\scriptscriptstyle\ EH-HYP1\ }$ & $(\Delta\bVstat)_{\rm\scriptscriptstyle\ EH-HYP2\ }$ \\
        \hline\\[-2.0ex]
        0.5 & 6.0219 & 0.2094(18) & 0.1987(19) & 0.2826(23) \\
            & 6.1628 & 0.2020(24) & 0.1983(25) & 0.2793(28) \\
            & 6.2885 & 0.2050(28) & 0.2018(30) & 0.2828(34) \\
            & 6.4956 & 0.1789(51) & 0.1777(52) & 0.2501(57) \\ \\[-2.0ex]
        \hline\hline
      \end{tabular}
      \vskip 0.3cm
      \begin{tabular}{ccccc}
        \hline\hline
        $\theta$ & $\beta$ & $(\Delta\bVstat)_{\rm\scriptscriptstyle APE-HYP1}$ & $(\Delta\bVstat)_{\rm\scriptscriptstyle APE-HYP2}$ & $(\Delta\bVstat)_{\rm\scriptscriptstyle HYP1-HYP2}$ \\
        \hline\\[-2.0ex]
        0.5 & 6.0219 & -0.0106(9)  & 0.0729(13) & 0.0835(7)  \\
            & 6.1628 & -0.0036(10) & 0.0771(14) & 0.0807(8)  \\
            & 6.2885 & -0.0032(11) & 0.0776(15) & 0.0808(8)  \\
            & 6.4956 & -0.0004(15) & 0.0720(21) & 0.0724(12) \\ \\[-2.0ex]
        \hline\hline
      \end{tabular}
    \end{center}
  \small
    \begin{center}
      \vbox{\vskip 0.6cm}
      \begin{tabular}{ccccc}
        \hline\hline
        $\theta$ & $\beta$ & $(\Delta\bVstat)_{\rm\scriptscriptstyle\ EH-APE\ }$  & $(\Delta\bVstat)_{\rm\scriptscriptstyle\ EH-HYP1\ }$ & $(\Delta\bVstat)_{\rm\scriptscriptstyle\ EH-HYP2\ }$ \\
        \hline\\[-2.0ex]
        1.0 & 6.0219 & 0.1332(25) & 0.1250(27) & 0.2019(34) \\
            & 6.1628 & 0.1301(30) & 0.1270(30) & 0.2018(34) \\
            & 6.2885 & 0.1391(34) & 0.1379(36) & 0.2153(41) \\
            & 6.4956 & 0.1284(63) & 0.1285(64) & 0.2013(68) \\ \\[-2.0ex]
        \hline\hline
      \end{tabular}
      \vskip 0.3cm
      \begin{tabular}{ccccc}
        \hline\hline
        $\theta$ & $\beta$ & $(\Delta\bVstat)_{\rm\scriptscriptstyle APE-HYP1}$ & $(\Delta\bVstat)_{\rm\scriptscriptstyle APE-HYP2}$ & $(\Delta\bVstat)_{\rm\scriptscriptstyle HYP1-HYP2}$ \\
        \hline\\[-2.0ex]
        1.0 & 6.0219 & -0.0082(12) & 0.0686(19) & 0.0768(10) \\
            & 6.1628 & -0.0030(12) & 0.0716(16) & 0.0747(8)  \\
            & 6.2885 & -0.0012(13) & 0.0761(18) & 0.0772(10) \\
            & 6.4956 &  0.0001(17) & 0.0728(24) & 0.0727(13) \\ \\[-2.0ex]
        \hline\hline
      \end{tabular}
    \end{center}
    \caption{Non-perturbative determinations of $\Delta\bVstat$ for various gauge couplings and static actions. Different choices of $\theta$
        correspond to independent improvement conditions.\label{tab:dbVstatnp}}
\end{table}

\vbox{

\end{document}